\documentclass[12pt,a4paper]{article}

\usepackage[british]{babel}

\usepackage[a4paper,top=2cm,bottom=2cm,left=2.5cm,right=2.5cm,marginparwidth=1.75cm]{geometry}

\usepackage[style=ieee, backend=biber, maxcitenames=1, mincitenames=1, maxbibnames=999, minbibnames=1]{biblatex}
\DeclareFieldFormat[article, inbook, incollection, inproceedings, misc, thesis, unpublished]{title}{#1}  
\DeclareFieldFormat{title}{#1}  
\DeclareFieldFormat{journaltitle}{#1}  
\DeclareFieldFormat{booktitle}{#1}  
\DeclareFieldFormat{maintitle}{#1}  
\DeclareFieldFormat{issuetitle}{#1}  
\DeclareFieldFormat{eventtitle}{#1}  

\DeclareFieldFormat{url}{\url{#1}}
\DeclareFieldFormat{doi}{\newline\mkbibacro{DOI}\addcolon\space\url{#1}}

\usepackage[table,xcdraw]{xcolor}
\usepackage{float}
\usepackage{makecell}
\usepackage{amsmath}
\usepackage{graphicx}
\usepackage[colorlinks=true, allcolors=blue]{hyperref}
\usepackage{hyperref}
\usepackage{orcidlink}
\usepackage[title]{appendix}
\usepackage{mathrsfs}
\usepackage{amsfonts}
\usepackage{booktabs} 
\usepackage{caption}  
\usepackage{threeparttable} 
\usepackage{algorithm}
\usepackage{algorithmicx}
\usepackage{algpseudocode}
\usepackage{listings}
\usepackage{enumitem}
\usepackage{chngcntr}
\usepackage{booktabs}
\usepackage{lipsum}
\usepackage{subcaption}
\usepackage{authblk}
\usepackage[T1]{fontenc}    
\usepackage{csquotes}       
\usepackage{diagbox}
\usepackage{tabularx}
\usepackage{multirow}
\usepackage{rotating}
\usepackage{bm}
\usepackage{wrapfig}

\usepackage{setspace}
\usepackage{titlesec}
\titleformat{\section} 
  {\normalfont\Large\bfseries}{\thesection.}{1em}{}
  
\usepackage{lineno} 

\rightlinenumbers 

\linenumbers 

\usepackage{float}   
\usepackage{caption} 


\newcolumntype{C}[1]{>{\centering\arraybackslash}m{#1}}

\title{An Uncertainty-aware Deep Learning Framework-based Robust Design Optimization of Metamaterial Units}

\author[1]{Zihan Wang}
\author[2,*]{Anindya Bhaduri}
\author[1]{Hongyi Xu}
\author[2]{Liping Wang}
\affil[1]{\small Mechanical Engineering, University of Connecticut, Storrs, CT 06269}
\affil[2]{Probabilistic Design, GE Aerospace Research, Niskayuna, NY, 12309}
\affil[*]{Corresponding author: \texttt{anindya.bhaduri@ge.com}}

\date{}  

\begin{document}
\maketitle
\begin{abstract}
Mechanical metamaterials represent an innovative class of artificial structures, distinguished by their extraordinary mechanical characteristics, which are beyond the scope of traditional natural materials. The use of deep generative models has become increasingly popular in the design of metamaterial units. The effectiveness of using deep generative models lies in their capacity to compress complex input data into a simplified, lower-dimensional latent space, while also enabling the creation of novel optimal designs through sampling within this space. 
However, the design process does not take into account the effect of model uncertainty due to data sparsity or the effect of input data uncertainty due to inherent randomness in the data. This might lead to the generation of undesirable structures with high sensitivity to the uncertainties in the system. To address this issue, a novel uncertainty-aware deep learning framework-based robust design approach is proposed for the design of metamaterial units with optimal target properties. The proposed approach utilizes the probabilistic nature of the deep learning framework and quantifies both aleatoric and epistemic uncertainties associated with surrogate-based design optimization. We demonstrate that the proposed design approach is capable of designing high-performance metamaterial units with high reliability. To showcase the effectiveness of the proposed design approach, a single-objective design optimization problem and a multi-objective design optimization problem are presented. The optimal robust designs obtained are validated by comparing them to the designs obtained from the topology optimization method as well as the designs obtained from a deterministic deep learning framework-based design optimization where none of the uncertainties in the system are explicitly considered.
\end{abstract}


\section{Introduction}\label{sec:intro}
Traditional materials are defined by their physical characteristics such as mechanical, electromagnetic, thermal, and optical behaviors, which stem from their molecular or atomic make-up. This composition can be manipulated to customize these properties for specific applications. Metamaterials, made up of individual units known as a "meta" cell, exhibit properties that rely on their unique spatial configuration. They achieve extraordinary characteristics through the precise arrangement of the "meta" cell. Essentially, any conventional material can be organized spatially into a unit that can be repetitively structured into a metamaterial. The design of metamaterial units is crucial for exploring and discovering new structures that possess exceptional mechanical properties, such as having unique stiffness-to-weight ratio \parencite{zheng2014ultralight}, capabilities in acoustic damping \parencite{chen2007acoustic}, capturing waves \parencite{wang2022design,wang2021gaussian,wang2022phononic,gurbuz2021generative}, reducing vibrations \parencite{qian2021optimization,claeys2017design,garland2020coulombic}, and absorbing energy efficiently \parencite{xu2019control,alberdi2020multi}, etc. Metamaterials hold promising potential for use in a wide range of areas \parencite{kumar2022overview}, including aerospace and seismic engineering, biomechanics and medical devices, sports equipment manufacturing, among others.

Deep learning (DL) has emerged as a powerful tool in computational metamaterial design, with extensive research highlighting its potential \parencite{yang2018porosity,liu2016materials,jha2018extracting,cang2017microstructure,wang2022design,wang2020data,meyer2022graph,bastek2022inverting,kumar2020inverse}. In particular, deep generative models like the variational autoencoder (VAE) and generative adversarial networks (GAN), along with their variants, have become prevalent for the inverse design of metamaterial units. VAEs, noted for their ability to generate a structured, continuous, and explicit low-dimensional design space and for their stable training process \parencite{wang2020deep,wang2024manufacturability,zheng2023unifying,wang2024generative,liu2020hybrid}, have gained popularity over GANs in metamaterial design applications. For example, \textcite{wang2020deep} introduced a VAE framework for creating functionally graded and heterogeneous metamaterial systems designed for specific distortion behaviors. \textcite{wang2022design} developed a Gaussian-Mixture VAE model for learning features of 2D metamaterial units and performing inverse design to achieve units with targeted mechanical properties. \textcite{zheng2023unifying} combined a VAE with a property predictor in a graph-based framework to optimize truss designs for desired mechanical properties in both linear and nonlinear domains. \textcite{wang2024generative} used a VAE to understand design-performance relationships, enabling the creation of graded mechanical metamaterial arrays with specified performance targets. \textcite{liu2020hybrid} applied a VAE for pixelated optical metasurface designs, utilizing evolutionary algorithms for optimization within the learned design space. These studies exemplify the use of VAEs for reconstructing latent feature spaces of metamaterials and performing inverse design to identify optimal configurations. 

However, applying deep generative models for inverse metamaterial design poses significant challenges. The effectiveness of these models heavily depends on the quality of the trained deep generative model. In many engineering scenarios, the dataset size may be limited, raising concerns that the training data might not adequately represent the entire design space \parencite{wang2022design}. This limitation risks creating models biased towards known data, potentially overlooking innovative or uncharted design areas. Optimizations in such constrained spaces introduce significant uncertainties, possibly leading to designs with imprecise property predictions. Broadly, uncertainty is categorized into two main types: epistemic and aleatoric. Epistemic uncertainty refers to the the lack of complete knowledge in the model (that characterizes the dataset of interest) parameters and it most often arises from insufficient training data. This type of uncertainty can potentially be reduced through the augmentation of training data size. Aleatoric uncertainty, conversely, is attributed to the intrinsic variability in the data that remains constant regardless of additional data collection. It is generally recognized that uncertainty is inevitable in engineering design. Therefore, it's essential to develop models that not only produce dependable designs but also precisely assess the uncertainties involved. Although numerous studies utilize deep generative models for metamaterial design, very few studies address the challenge of quantifying the associated uncertainties. Chen et al. \parencite{chen2023gan} explored hierarchical deep generative models for generating metasurfaces with geometric uncertainty, offering a compact representation of ideal designs and their conditional distributions. \textcite{yang2021general} proposed a general framework that combines a generative adversarial network and a mixture density network for microstructural material design, and has been shown to produce multiple promising solutions. Nevertheless, these approaches focus solely on geometrical designs and fall short of property-driven designs. To the best of the authors’ knowledge, there exists no other research that focuses on quantifying the uncertainty of the metamaterial designs obtained by deep generative models, which underscores an area ripe for further investigation.

Probabilistic deep learning models represent a significant advancement in the field of artificial intelligence, offering a framework to capture and express uncertainty in predictions and inferences \parencite{chang2021probabilistic,durr2020probabilistic,deshpande2022probabilistic, mufti2024shock}. Probabilistic deep learning models have been widely used in the application of path planning and decision making \parencite{wang2021robot}, disease diagnosis and drug discovery \parencite{hirschauer2015computer,er2012approach,mantzaris2011genetic,niwa2004prediction}, robotics navigation \parencite{karakucs2013learning,nguyen2023uncertainty}, forecast product demand \parencite{parry2011forecasting}, etc. One major class of probabilistic deep learning models are probabilistic deep neural networks (PDNNs) \parencite{specht1990probabilistic, ravi2023probabilistic}, which relies on the integration of probabilistic layers in the deep neural networks. Mixture density network (MDN) \parencite{bishop1994mixture} is one of the widely used approaches. This method models the final output as a distribution of possible values rather than a single deterministic value as with typical neural networks or other surrogate models \cite{williams1998prediction, cristianini2000introduction, bhaduri2018efficient, bhaduri2020free, ravi2023uncertainty}. In the area of design of metamaterials, \textcite{unni2021mixture} proposed a deep convolutional mixture density network for the inverse design of photonic structures, which models the design parameters as a multimodal probability distribution, which gives valuable information about the uncertainty in prediction.  \textcite{yang2021general} proposed a general framework that combines a GAN and an MDN for inverse modeling in microstructural material design. The findings from their study indicate that this integrated approach is capable of generating several viable solutions. \textcite{unni2020deep} proposed a tandem optimization model that combines an MDN and a fully connected network to inverse design practical thin-film high reflectors. The proposed model combines the high-efficiency advantages of DL with the optimization-enabled performance improvement, enabling efficient inverse design. Apart from Probabilistic Deep Neural Networks (PDNNs), deep generative models (DGMs) inherently possess the capability to quantify uncertainty \parencite{bohm2019uncertainty,sensoy2020uncertainty}. These models incorporate probabilistic approaches within their architecture, allowing them to represent and quantify the uncertainty in their predictions or generated outputs. As such, probabilistic deep learning models open new avenues for advancing the design and optimization of metamaterials, enabling the exploration of previously inaccessible design territories with a greater degree of confidence and risk management.

Prior deep generative model-based design methodologies have not thoroughly accounted for uncertainties inherent in the deep generative models. The objective of this work is to propose an uncertainty-aware deep generative model-based approach for the robust design of metamaterial units. First, an uncertainty-aware deep learning framework is proposed, which combines a VAE and an MDN network for modeling both the geometry of the metamaterial units and their corresponding mechanical properties by probability distributions. After training the proposed deep learning framework, we propose a deep learning framework-based robust design optimization that leverages the probabilistic nature of the VAE and the MDN networks to capture both aleatoric and epistemic uncertainties. This design approach aims to generate 3D metamaterial units for optimal properties with reduced sensitivity to the associated uncertainties. Our contribution of this work is threefold: 
\begin{itemize}
\item We present an uncertainty-aware deep learning framework tailored for metamaterial units, with an emphasis on quantifying both aleatoric and epistemic uncertainty.
\item We propose a progressive transfer learning-based training strategy that enhances model convergence and efficiency. This approach is instrumental in optimizing the balance between different loss terms, demonstrating its efficacy in fine-tuning the model for superior performance. 
\item Leveraging the uncertainty-aware deep learning framework, we propose a design methodology for creating robust metamaterial units. This approach incorporates uncertainty into the design process, ensuring the generated designs are not only innovative but also reliable and resilient to any variability in the system.
\end{itemize}

The remaining of the paper is organized as follows: Section \ref{sec:methodology} presents our proposed design approach for designing robust metamaterial units using an uncertainty-aware deep learning framework, along with an analysis of the uncertainty sources within the model. We also proposed a progressive transfer learning-based training strategy for the model training. In Section \ref{sec:data}, the data generation process is discussed. In Section \ref{sec:result}, we show the training and validation results of the proposed deep learning framework. Additionally, we validate the uncertainty-aware deep generative model-based design approach by two robust design cases. Conclusions are made in Section \ref{sec:conclusion}.


\section{Methodology}\label{sec:methodology}

The overarching goal of this design approach is to quantify both aleatoric and epistemic uncertainty in the deep generative model and, therefore perform inverse robust design to find the metamaterial unit’s configuration that corresponds to the desired mechanical properties. The proposed design approach consists of two parts:

(1) Training a deep learning framework to predict properties under uncertainty given the 3D metamaterial architecture and also obtain an intermediate low-dimensional latent feature space: This model comprises two key components - a DGM for learning low-dimensional features and a PDNN for predicting properties. The predictions from the PDNN include both mean values and standard deviations, providing a probabilistic understanding of the mechanical property behavior. To enhance model training, we also introduce a progressive transfer learning-based strategy. Further details about this uncertainty-aware deep learning framework are provided in Section \ref{subsec:generativemodel}.

(2) Performing robust design optimization on the trained latent feature space: The trained latent feature space is utilized to design novel metamaterial units, taking into account both aleatoric and epistemic uncertainty. The robust design optimization is carried out using the NSGA-II algorithm \parencite{blank2020pymoo}, a multi-objective evolutionary algorithm known for its effectiveness in avoiding local optima and reaching global optima. The optimization aims to minimize the combined influence of the predicted mean and the associated uncertainty (standard deviation) of various property values, ensuring the design meets multiple performance criteria simultaneously while adhering to necessary constraints. The measurement of the predicted mean and the associated uncertainty are explained in section \ref{subsec:uncertainty}.

\subsection{Probabilistic deep learning framework}\label{subsec:generativemodel}

Probabilistic deep learning is a branch of deep learning designed to address uncertainty. There are two key methodologies within probabilistic deep learning: probabilistic deep neural networks (PDNN) and deep generative models (DGM). In PDNNs, deep neural networks integrate probabilistic layers or elements to effectively manage and model uncertainty, while DGMs fuse probabilistic models with deep neural network elements to capture intricate, nonlinear stochastic connections among random variables.  

The proposed uncertainty-aware deep generative model consists of two parts: 

(1) A DGM employing 3D convolutional layers in both the encoder and decoder to map high-dimensional input 3D structures into a probabilistic lower-dimensional latent space.

(2) A PDNN mapping the mean features of the latent space to the mean and standard deviation of the mechanical properties.

In this work, we opt for VAE as the deep generative model, and MDN as the probabilistic deep neural network. Nonetheless, these models can be readily substituted with other types of PDNNs and DGMs within the framework of the overall design approach.

\subsubsection{Probabilistic Deep Neural Networks (PDNNs)}
PDNNs are specialized neural networks enhanced with probabilistic layers or elements, designed specifically to address and manage uncertainty within their architecture. These networks are adapted from conventional neural network structures to better capture the nuances of uncertainty in data and predictions. Broadly categorized into two types, PDNNs are employed for their unique approaches to quantifying uncertainty. The first type leverages statistical methods to fine-tune parameters, optimizing for the observed data's probability distribution rather than settling for mere point estimates. Within this category, Quantile Regressions (QRs) \parencite{koenker2005quantile} and Mixture Density Networks (MDNs) \parencite{specht1990probabilistic} are particularly notable for their effectiveness. The second type of PDNNs incorporates explicit probabilistic layers aimed to capture model uncertainty, with Bayesian Neural Networks (BNNs) \parencite{kononenko1989bayesian}, Monte Carlo Dropout (MC Dropout) \parencite{milanes2021monte}, and Deep Ensemble Learning (DELs) \parencite{ganaie2022ensemble} being prominent examples. 

In this study, our primary focus is on MDN, a specialized form of neural network designed to solve inverse problems. Unlike traditional neural networks that predict a singular output value, MDN aims to forecast the entire probability distribution of the output given an input. Specifically, an MDN typically employs a neural network to parameterize a mixture model, which is often comprised of several predefined distributions. Generally, Gaussian distribution is used, and the output is modeled as a conditional probability \(P(y \mid z)\), expressed as:
\begin{equation}
P(\bm{y}|\bm{z}) = \sum_{k=1}^{K} \pi_k(\bm{z}) \mathcal{N}(\bm{z} | \{\mu_k(\bm{z}), \sigma_k(\bm{z})\}), \quad \sum_{k=1}^{K} \pi_k(\bm{z}) = 1
\end{equation}
where $K$ is the total number of individual Gaussian distributions, $\bm{z}$ and $\bm{y}$ are the inputs and outputs of the network, respectively, \(\pi_k\) represents the mixing coefficients, \(\mu_k\) and \(\sigma_k\) are the mean and standard deviation of the \(k\)th Gaussian distribution, respectively. To optimize the network, the goal is to minimize the negative log-likelihood of the predicted distribution against the training data:
\begin{equation}\label{eq:MDNloss}
L_{\text{MDN}}= -\frac{1}{N} \sum_{n=1}^{N} \log \left( \sum_k \pi_k(\bm{y}_n | {\mu}_k(\bm{z}_n; w), {\sigma}_k(\bm{z}_n; w)) \right)
\end{equation}
where \(N\) is the batch size, \(w\) are the weights in the MDN network, $\bm{z_n}$ is the \(n\)th instance in a batch, and $\bm{y_n}$ is the corresponding label. This approach highlights the MDN's ability to capture intricate probabilistic input-output relationships, providing a more detailed and insightful prediction model than traditional neural networks. In our work, we simplify the MDN by setting \(k=1\) in our MDN network, therefore, the MDN model parameterizes a single Gaussian distribution.

\subsubsection{Deep Generative Models (DGMs)}
DGMs are neural networks trained to approximate complicated, high-dimensional probability distributions using samples. When trained successfully, we can use the DGM to estimate the likelihood of each observation and to create new samples from the underlying distribution. DGMs include generative adversarial networks (GANs) \parencite{zheng2021controllable,lai2021conditional,jin2022intelligent}, variational autoencoders (VAEs) \parencite{liu2020hybrid,wang2024generative,zheng2023unifying,wang2022design}, diffusion models \parencite{bastek2023inverse,zhang2023diffusion}, etc. Among these models, we specifically chose to employ a VAE for its training stability, explicit representation of latent space and efficient inference.

VAE, originated from the autoencoder and contains two components: an encoder and a decoder. The VAE’s encoder conducts nonlinear dimensionality reduction and compresses the high-dimensional data $\bm{x}$ into a low-dimensional latent space $\bm{z}$. The encoder can be expressed as \(Q_\phi(\bm{z} \mid \bm{x})\), which is the approximate posterior that follows a normal distribution, where \(\phi\) is the vector of the encoder parameters. The decoder, also a nonlinear operator, can map back the low-dimensional latent feature space to the original high-dimensional input data space. The decoder is expressed as \(P_\theta(\bm{x} \mid \bm{z})\), where \(\theta\) is the vector of decoder parameters. The VAE integrates Bayesian inference with the autoencoder architecture, encouraging regularization of the latent feature space towards a Gaussian distribution. This process introduces a measure of variability in the latent space, which reflects the model uncertainty about the latent representations of the given dataset. 
In this paper, we recognize the uncertainty caused by the latent space as latent space uncertainty.

The loss function of VAE includes two parts, and it can be expressed as:
\begin{equation}\label{eq:VAEloss}
L_{\text{VAE}} = L(\bm{x}, \hat{\bm{x}}) + L_{\text{KL}}(\bm{z}, \mathcal{N}(0, I_d))
\end{equation}
where \(\bm{z}\) represents the latent vectors, \(\bm{x}\) represents the input data, and \(\hat{\bm{x}}\) represents the reconstruction data. \(L(\bm{x}, \hat{\bm{x}})\) is the mean squared reconstruction loss between \(\hat{\bm{x}}\) and \(\bm{x}\), defined by \(L(\bm{x}, \hat{\bm{x}}) = \frac{1}{n} \sum_{i=1}^{n} (\bm{x} - \hat{\bm{x}})^2\), where \(n\) represents the number of training data in the VAE model. \(L_{\text{KL}}(\bm{z}, \mathcal{N}(0, I_d))\) is the Kullback-Leibler divergence loss, which measures the differences between the distribution of latent vector \(\bm{z}\) and the standard normal distribution \(\mathcal{N}(0, I_d)\).

\subsubsection{Proposed deep learning framework}\label{subsubsec:DLframe}
\begin{figure}[!ht]
\centering
\includegraphics[width=1\linewidth]{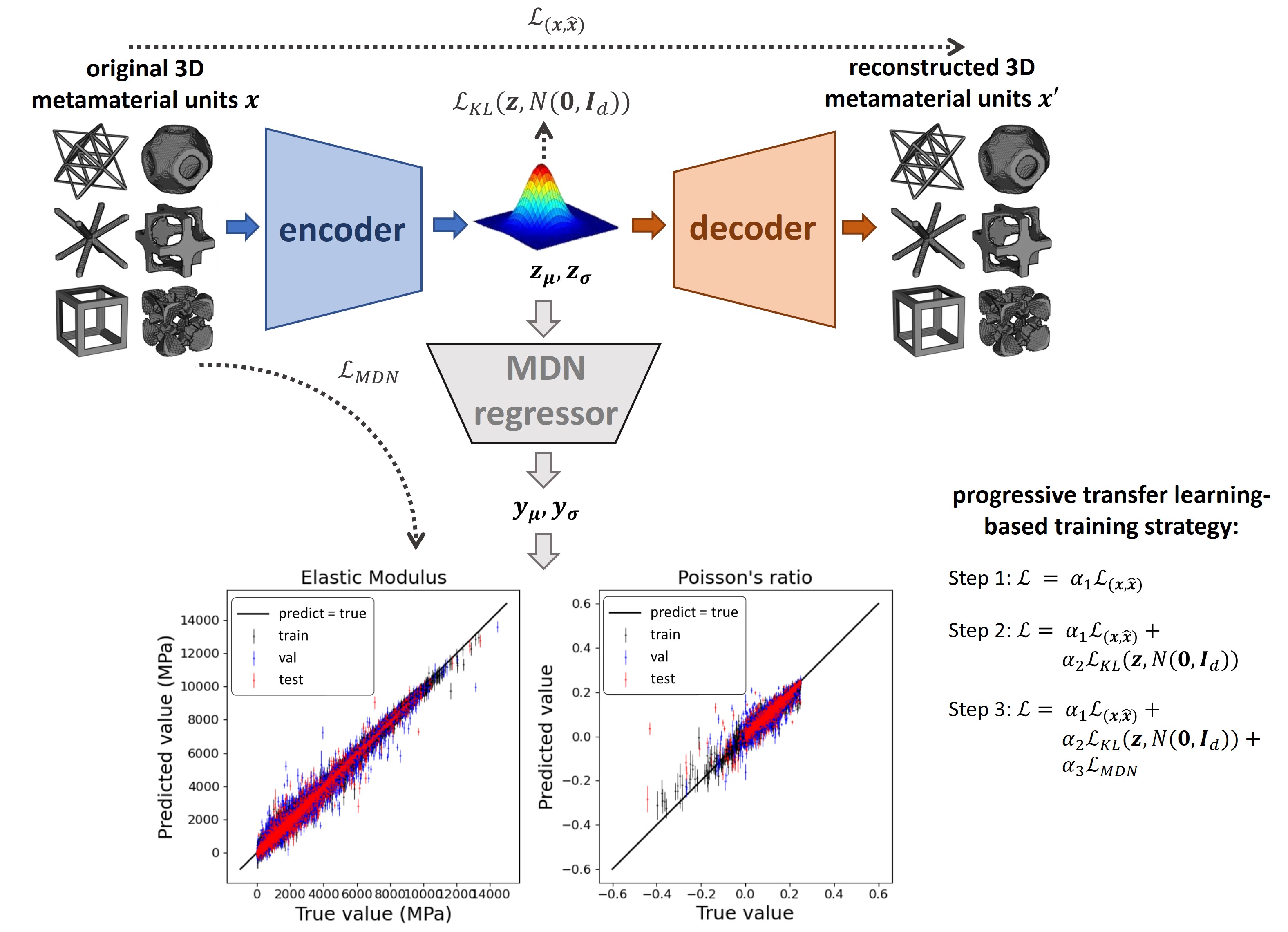}
\caption{\label{fig:vae}An uncertainty-aware deep learning framework is employed to characterize 3D metamaterial units and their mechanical properties, incorporating uncertainty in the analysis. This model is composed of two primary elements: a DGM that extracts low-dimensional features and a PDNN that forecasts properties. The outputs of the PDNN, encompassing both mean values and standard deviations, offer a probabilistic interpretation of the mechanical behaviors.}
\end{figure}

Research \parencite{xu2022harnessing,wang2020deep} extensively validates that integrating the property predictor with the latent space of the VAE model effectively captures the relationships between structure and properties. In light of these findings, we have developed a model that enhances the VAE by integrating a Mixture Density Network (MDN) into its latent feature space (Figure \ref{fig:vae}). The hyperparameters of the uncertainty-aware deep learning framework are shown in Table \ref{tab:Deepgenerativeparam} in Appendix \ref{sec:appendix2}. Both the VAE and MDN components of the model are trained simultaneously. The combined loss function for this training process aggregates the loss terms from both the VAE (Equation \ref{eq:VAEloss}) and the MDN (Equation \ref{eq:MDNloss}), expressed as follows: 
\begin{equation}\label{eq:losses}
L_{\text{all}} = L_{\text{VAE}} + L_{\text{MDN}} = \alpha_1 L(\bm{x}, \hat{\bm{x}}) + \alpha_2 L_{\text{KL}}(\bm{z}, \mathcal{N}(0, I_d)) + \alpha_3 L_{\text{MDN}}
\end{equation}
where $\alpha_1$, $\alpha_2$, $\alpha_3$ represent the coefficients among different loss terms. In order to balance different loss terms and find the best combination of these coefficients, we propose a progressive transfer learning-based training strategy, which is shown in section \ref{subsec:transfer}.

Figure \ref{fig:vae} presents our model, which is designed to analyze 3D metamaterial units and predict their mechanical properties under uncertainty. The model provides outputs that include both mean values and standard deviations for a probabilistic interpretation of mechanical behaviors. In our model, we specifically address two categories of uncertainty:

(1) Data Uncertainty: This type of uncertainty, categorized as aleatoric uncertainty, arises from the inherent imprecision and variability present in the input mechanical properties and the 3D metamaterial structures themselves. Our model is adept at quantifying this uncertainty, capturing both the inherent fluctuations in mechanical properties and the diversity in structural configurations.

(2)	Latent Space Uncertainty: This type of uncertainty, known as epistemic uncertainty, refers to the variability encountered in the process of reconstructing samples from the latent space, as well as the intrinsic variability of the generated samples themselves. This uncertainty underscores the challenges in accurately replicating the input data or generating new, realistic samples based on trained distributions.

\subsubsection{Progressive Transfer Learning-based Training Strategy}\label{subsec:transfer}
To determine the optimal combination of coefficients for each loss term in Equation \ref{eq:losses}, we propose a progressive transfer learning-based training strategy to enhance the training of the deep learning framework. The core concept of this strategy is to identify the ideal dimensionality of the latent feature space and progressively adjust each loss term to achieve the best model convergence. Our training strategy is outlined as follows:

\begin{itemize}
\item Step 1: In the development of our model, achieving high reconstruction accuracy of the metamaterial units is most important. Thus, we initially set \(\alpha_1 = 1\) and temporarily set \(\alpha_2 = \alpha_3 = 0\) to determine the optimal dimensionality of the latent feature space. While a larger latent space dimension can improve reconstruction accuracy, it also increases the computational demands, particularly during design optimization processes on the latent feature space. Therefore, we implement a comparative analysis to select the dimensions of the latent feature space, starting from a minimal dimensionality and progressively increasing until achieving satisfactory reconstruction accuracy.
 
\item Step 2: In the second step of the methodology, we set \(\alpha_1=1\), \(\alpha_3=0\), and proceed to incrementally change the \(\alpha_2\) value. The model weights pre-trained in Step 1 serve as the initial weights for subsequent training phases. With each increase in \(\alpha_2\), we utilize the optimally trained weights from the preceding iterations as the initial values for the next phase of model training. This approach ensures a smooth and informed transition between training phases, leveraging accumulated learning to refine the model's performance progressively. For each phase of training, the reconstruction accuracy and the KL divergence loss are monitored and recorded. The best \(\alpha_2\) value is identified by the best reconstruction accuracy as well as the lowest KL divergence loss.

\item Step 3: In this step, we use the \(\alpha_1\) and \(\alpha_2\) values determined in the previous step and incrementally change the \(\alpha_3\) value. The model weights pre-trained in Step 2 serve as the initial weights for this phase of training. With each increment of \(\alpha_3\), the weights from the preceding phase are used as the starting point for the next phase. The optimal \(\alpha_3\) value is identified when the model achieves the best balance between reconstruction error, KL divergence loss, and regression error.
\end{itemize}

\subsection{Robust design optimization}\label{subsec:uncertainty}

Design under uncertainty has been gaining attention for decades, which aims to account for stochastic variations in engineering design (e.g., material, geometry, property, condition). Many approaches in literature incorporated uncertainty into a design formulation. Robust design optimization, first proposed by \textcite{tsui1992overview}, seeks to mitigate the effects caused by variations without actually removing these causes. Reliability-based design \parencite{choi2007reliability} incorporates reliability engineering principles into the design process, which focuses on ensuring that the product or system performs its intended function under stated conditions over time. Probabilistic design \parencite{long1999probabilistic} employs probability theory to account for uncertainties in design parameters and environmental conditions. 

In this work, we focus on the robust design optimization and the goal is to obtain optimal structures under uncertainty when the values of certain properties of interest are maximized. The design approach is thus stated as:
\begin{align}\label{eq:robust}
\max_{\bm{z}} &\left[ \mu(f_1(\bm{z})) - \beta_1\sigma(f_1(\bm{z})), \mu(f_2(\bm{z})) - \beta_2\sigma(f_2(\bm{z})), \ldots, \mu(f_{n_f}(\bm{z})) - \beta_n\sigma(f_{n_f}(\bm{z})) \right] \nonumber \\
\text{s.t.} \quad &c_j(\bm{z}) \leq 0
\end{align}
where \(\bm{z}\) is a vector of design variables in the form of the latent variable vector learned from the deep learning framework. $\mu(f_i(\bm{z}))$ and $\sigma(f_i(\bm{z}))$ are the mean and standard deviation of the predicted property values $f_i(\bm{z})$, respectively, where \(i=1,2,\ldots,n_f\) and \(n_f\) is the number of property values. $\beta_i$ is the weight parameter that modulates the importance of the mean compared to the standard deviation, which can vary across different property values. \(c_j\) (\(j=1,2,\ldots,n_c\)) are the \(n_c\) number of constraint functions.

\begin{figure}[!ht]
\centering
\includegraphics[width=1\linewidth]{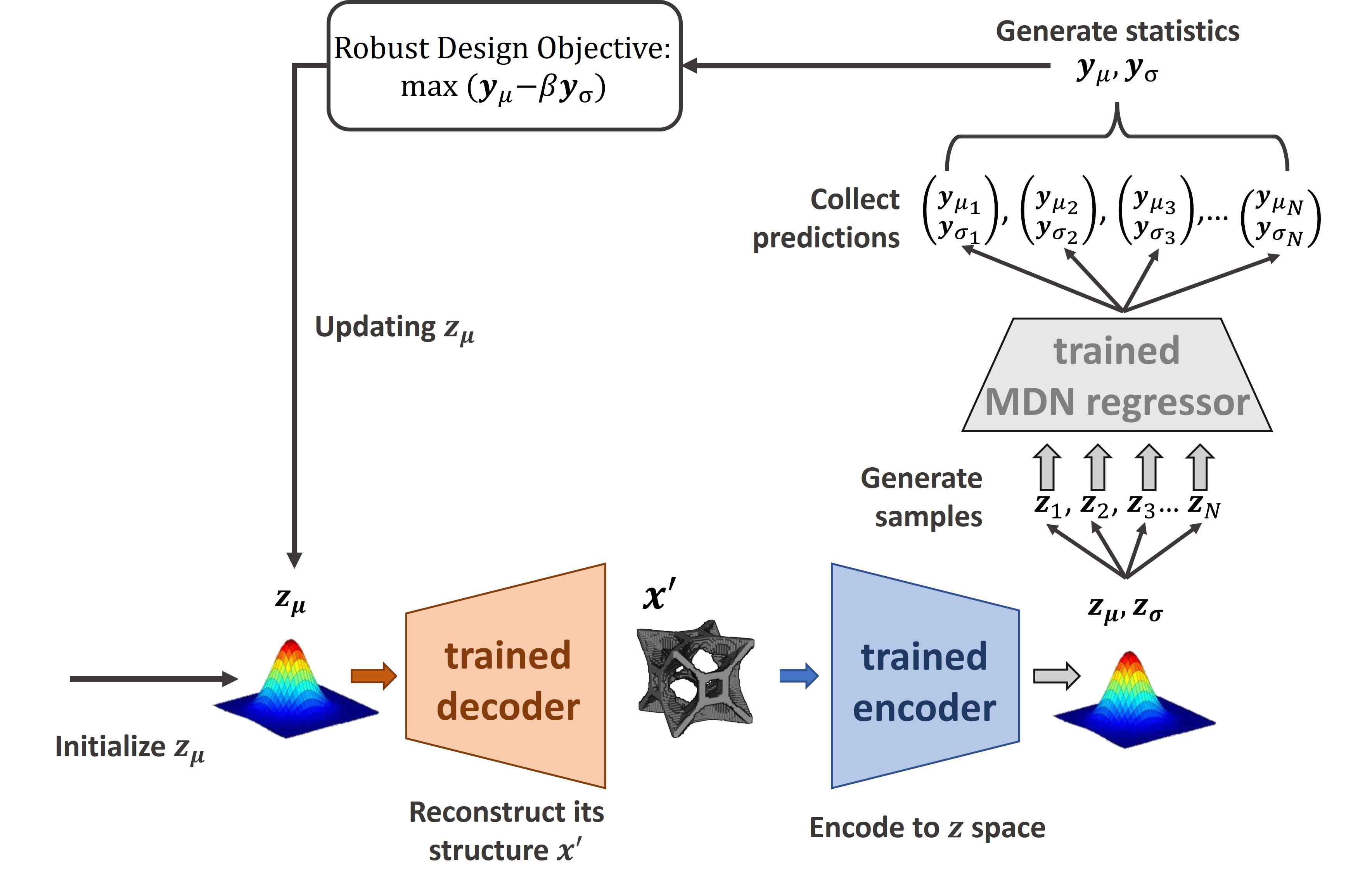}
\caption{\label{fig:framework} Performing robust design optimization on the trained latent feature space. The optimization seeks to reduce the aggregate impact of the predicted mean and corresponding uncertainty across various property values.}
\end{figure}

The uncertainty sources mentioned in the section \ref{subsubsec:DLframe} contribute to both aleatoric and epistemic uncertainties. Understanding and quantifying these uncertainties is crucial for robust design optimization, which necessitates both the mean value and the total uncertainty of the predicted mechanical properties. The process of quantifying these predictions involves several key steps in the latent feature space (Figure \ref{fig:framework}):

\begin{itemize}

\item Initialize \(\bm{z}_{\mu}\): After training the deep learning framework, the datasets are encoded into their corresponding latent vectors (\(\bm{z}_{\mu}\) and \(\bm{z}_{\sigma}\)). We choose a \(\bm{z}_{\mu}\) value as the optimization starting point.

\item Reconstruct its structure \(\bm{x}'\): Reconstruct the latent vectors' structure, denoted as \(\bm{x}'\).

\item Encode to \(\bm{z}\) space: The structure \(\bm{x}'\) is then re-encoded to determine their mean (\(\bm{z}_{\mu}\)) and standard deviation (\(\bm{z}_{\sigma}\)) in the latent space, encapsulating the inherent uncertainty of the model.

\item Generate samples: Sample from the Gaussian distribution with \(\bm{z}_{\mu}\) and \(\bm{z}_{\sigma}\) generates multiple latent vector values, \((\bm{z}_1, \bm{z}_2, \bm{z}_3, \ldots, \bm{z}_N)\), where \(N\) represents the total number of sampling points. A sufficient number of sampling points will effectively explore the space of possible designs. The determination of the number of \(N\) is illustrated in appendix \ref{sec:appendix3}.

\item Collect predictions: Each sample point \(\bm{z}_i\) within the latent space is associated with specific mechanical property predictions using MDN, given by a mean (\(\bm{y}_{\mu_i}\)) and a standard deviation (\(\bm{y}_{\sigma_i}\)). This leads to a collection of predicted property distributions \((\bm{y}_{\mu_1}, \bm{y}_{\mu_2}, \bm{y}_{\mu_3}, \ldots, \bm{y}_{\mu_N}\) and \(\bm{y}_{\sigma_1}, \bm{y}_{\sigma_2}, \bm{y}_{\sigma_3}, \ldots, \bm{y}_{\sigma_N})\).

\item Generate statistics: The aggregation of these predictions provides an overall mean (\(\bm{y}_{\mu}\)) and standard deviation (\(\bm{y}_{\sigma}\)) for the sampled designs, reflecting the expected performance and overall uncertainty \parencite{egele2022autodeuq}. 

\item Updating \(\bm{z}_{\mu}\): Updating  \(\bm{z}_{\mu}\) by solving Equation \ref{eq:robust}.

\end{itemize}

In the statistics generation step, the predictive mean and total uncertainty required for solving Equation \ref{eq:robust} are computed as follows:

\begin{equation}\label{eq:cal_mean}
\bm{y}_{\mu} = \mu(f(\bm{z})) = \frac{1}{N} \sum_{i=1}^{N} \bm{y}_{\mu_i}
\end{equation}

The aleatoric uncertainty and epistemic uncertainty can be expressed as:

\begin{equation}\label{eq:cal_alea}
\sigma_{\text{aleatoric}} = \frac{1}{N} \sum_{i=1}^{N} \bm{y}_{\sigma_i}
\end{equation}

\begin{equation}\label{eq:cal_epis}
\sigma_{\text{epistemic}} = \sqrt{\frac{1}{N-1} \sum_{i=1}^{N} \left(\bm{y}_{\mu_i} - \bar{\bm{y}}_{\mu} \right)^2}
\end{equation}
where \(\bar{\bm{y}}_{\mu} = \frac{1}{N} \sum_{i=1}^{N} \bm{y}_{\mu_i}\) is the mean of the overall mean values. The total uncertainty can be calculated by:

\begin{equation}\label{eq:cal_total_uncertainty}
\sigma_{\text{total}}(f(\bm{z})) = \bm{y}_{\sigma} = \sqrt{\sigma_{\text{aleatoric}}^2 + \sigma_{\text{epistemic}}^2}
\end{equation}

\section{Data Generation}\label{sec:data}

We developed a database that contains 46840 samples of metamaterial units. These 3D metamaterial units were generated or sourced using three distinct methods or resources. Each sample in the database has a resolution of 48x48x48 voxels.

\begin{figure}[!ht]
\centering
\includegraphics[width=1\linewidth]{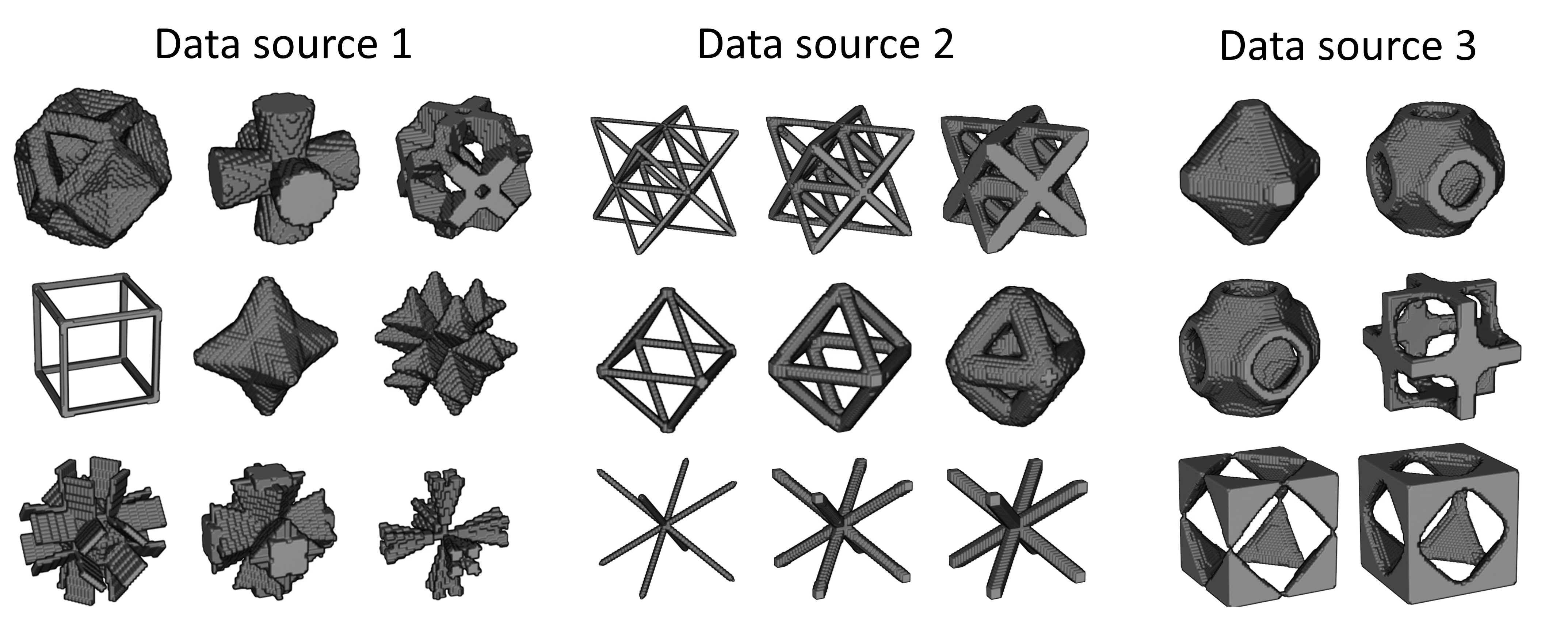}
\caption{\label{fig:dataset} Examples of metamaterial units in three data sources.}
\end{figure}

The first data source of 3D metamaterial units is generated using the microstructure family template-based method, modified from the one proposed in literature \parencite{chen2018computational}. The second data source comprises octet \parencite{deshpande2001effective}, octahedral \parencite{weiss1985geometry}, and body-centered cubic structures \parencite{mughbrabi1979persistent}. These are created by first outlining the skeleton of cubic symmetric metamaterial units within a continuous design space, and then forming the geometries by applying a radius along the outlined skeleton. The last source of 3D metamaterial units is collected from the open source dataset \parencite{chan2021metaset}, which is generated using level-set functions and creates isosurface families based on crystallographic structure factors. In all these three metamaterial unit generation methods/sources, we only generate/select the cubic symmetric metamaterial units with volume fraction in the range of [0.05,0.4] that leads to 46840 units. Examples of these metamaterial units are shown in Figure \ref{fig:dataset}. Detailed information about the generation and collection of the metamaterial units can be referred to our previous work \parencite{wang2024manufacturability}. Due to the significant variety in structural features and the unique aspects of the generation algorithms, it is impractical to capture the entirety of metamaterial unit samples using a few geometric parameters.

In this research, we aim to showcase our design approach by focusing on the exploration of the elasticity of metamaterial units. Aluminum has been selected as the base material due to its well-defined elastic properties, characterized by an elastic modulus \(E_{\text{Al}_{0}} = 68,300 \, \text{MPa}\) and a Poisson's ratio \(\nu_{\text{Al}_{0}} = 0.3\). 
To incorporate aleatoric uncertainty, which reflects the variability in input material properties, into our analysis, we adopt a probabilistic sampling approach for the elastic properties used in each simulation. Specifically, the values of \(E_{\text{Al}_{0}}\) and \(\nu_{\text{Al}_{0}}\) for each simulation sample are drawn from a Gaussian distribution with mean values \(\mathbf{\mu}\) set at their defined material property values (\(E_{\text{Al}}\) and \(\nu_{\text{Al}}\)), with corresponding standard deviations \(\sigma\) set as \(\sigma = 0.01 \mu\). This methodological choice enables us to systematically account for the inherent uncertainty in material
properties, ensuring that our simulation dataset comprehensively represents the potential variability in the elastic behavior of aluminum-based metamaterial units.
This framework has the potential to be extended to other base materials such as steel, titanium, copper, and Inconel. However, for each new material, simulations of the metamaterial unit properties will need to be re-conducted using the newly defined material properties.
The linear elastic properties of 3D metamaterial units are simulated using a user-defined linear elastic analysis subroutine in ABAQUS, along with unified Periodic Boundary Conditions (PBC) \parencite{xia2003unified}. In this work, the boundary conditions apply constant deformation to two opposing faces of the samples, focusing primarily on elastic deformation. Under steady-state conditions, stress and strain within the volume of the metamaterial units adhere to Hooke's Law. The resulting stress and strain data from these simulations allow for the calculation of the effective Young’s modulus \(E\) for each sample. These moduli can be computed based on the recorded stress and strain values across the material.

The effective Young's modulus \(E\) and shear modulus \(G\) can be computed as follows:

\begin{equation}
E = \frac{1}{n} \sum_{i=1}^{n} \frac{\sigma_i}{\epsilon_i}
\end{equation}

\begin{equation}
G = \frac{1}{n} \sum_{i=1}^{n} \frac{\tau_i}{\gamma_i}
\end{equation}
where \(n\) represents the number of nodes where stress, strain, and shear are recorded. \(\sigma_i\) and \(\tau_i\) are the normal and shear stresses at the \(i\)th node, and \(\epsilon_i\) and \(\gamma_i\) are the corresponding strains. The Poisson’s ratio \(\nu\) is derived from the relationship between \(\sigma\), \(\epsilon\), and \(\gamma\) across the samples.

\begin{figure}[!ht]
\centering
\includegraphics[width=1\linewidth]{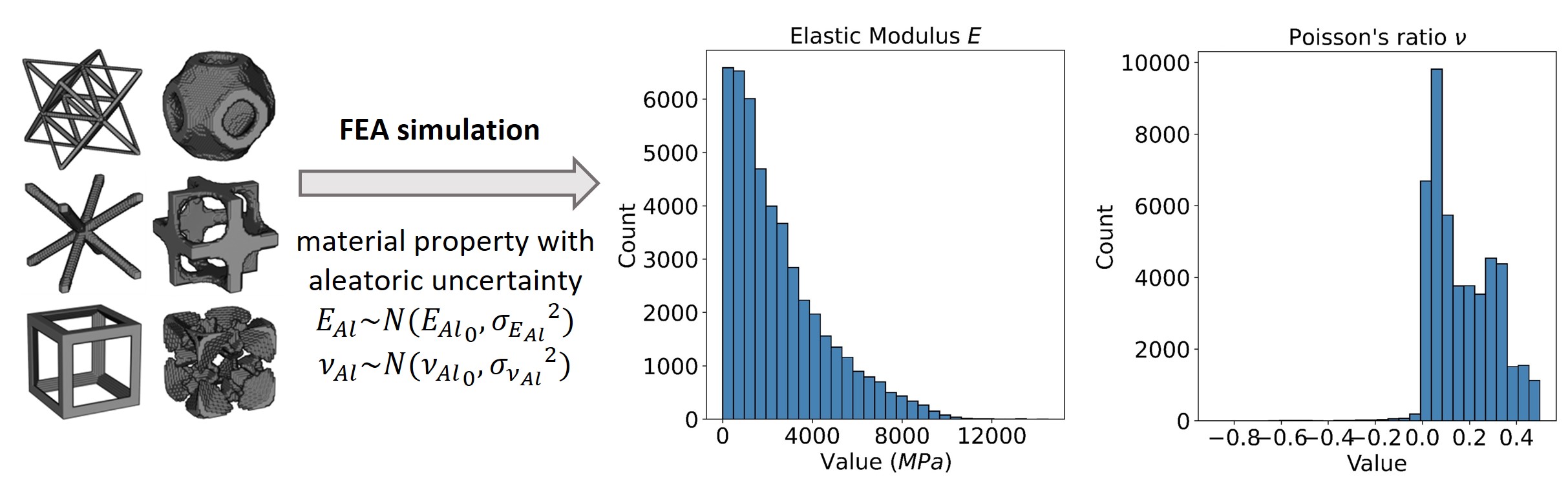}
\caption{\label{fig:dataset_hist} Generation process of the dataset and histograms of the elastic modulus and Poisson's ratio across all three data sources.}
\end{figure}

The simulations are conducted on all 46840 samples. The generation process and the histograms of elastic modulus and Poisson's ratio are displayed (Figure \ref{fig:dataset_hist}). The histograms offer insights into the range and variability of the elastic modulus and Poisson's ratio across all the metamaterial units in the dataset.

\section{Results}\label{sec:result}
\subsection{Structure-to-Property Mapping}

In this section, we first outline the training process of the deep learning framework, which is designed to map structures to their properties. Next, we validate the effectiveness of the trained model using several performance assessment metrics. Finally, we demonstrate the model's capability to generate new metamaterial samples by interpolating within the latent feature space.

\subsubsection{Training results}\label{subsubsec:training}

The 3D metamaterial unit dataset is divided into three sets, \(32788\) (\(70\%\)) for training, \(9368\) (\(20\%\)) for validation, and \(4684\) (\(10\%\)) for testing. To reduce the computational demands associated with training the deep learning framework, we exploit the inherent geometrical symmetry present in the metamaterial designs. By doing so, we utilize only an eighth of the entire structure for input, resulting in an input dimensionality of \(24 \times 24 \times 24\) voxels. To reconstruct the full structure of \(48 \times 48 \times 48\) voxels, the structures are mirrored three times. 

The proposed deep learning framework is implemented in PyTorch \parencite{paszke2019pytorch}. Adam is used as the optimizer for parameter optimization. The total number of epochs is set to \(400\). The initial learning rate is set to be \(0.001\) across all the models. To enhance the model's convergence towards optimal performance, an exponential decay strategy is employed, with a decay rate set at \(0.995\). Additionally, to prevent overfitting and unnecessary computation, an early stopping mechanism is integrated into the training process. This criterion halts the training if the validation loss fails to show improvement over \(10\) consecutive epochs. We use the proposed progressive transfer learning-based training strategy (discussed in section \ref{subsec:transfer}) to train the uncertainty-aware deep learning framework. The optimal dimensionality of the latent space was established through a parametric study, the results of which are detailed in Table \ref{tab:convergence_study} in Appendix. 
We found that a latent space dimensionality of 32 strikes the best balance between maintaining manageable dimensionality and achieving high reconstruction accuracy. This dimensionality was selected for its consistent performance without significantly increasing the complexity of the latent space. This decision was substantiated by comparing the relative errors for different dimensionalities, particularly noting minimal gains in accuracy beyond a dimensionality of 32. The process for identifying the optimal coefficients for the model's loss terms is illustrated in Tables \ref{tab:kl_loss_weights} and \ref{tab:reg_loss_weights} in Appendix. The coefficients \(\alpha_1 = 1\), \(\alpha_2 = 0.001\), \(\alpha_3 = 0.001\) were determined to be optimal based on achieving a balance between minimizing the KL divergence and the regression error while maximizing the reconstruction accuracy. These values facilitated effective learning of the model's underlying data structure, minimizing both overfitting and underfitting. This is evidenced by the improved loss metrics recorded during the training phases.

To demonstrate the advantages of our proposed progressive transfer learning-based training strategy, we conducted a comparative analysis between a model fine-tuned through progressive transfer learning and another model trained from scratch. Both models started with the same coefficient of loss terms (\(\alpha_1 = 1\), \(\alpha_2 = 0.001\), \(\alpha_3 = 0.001\)). The model developed from scratch showed significantly higher final loss values on the validation set,
highlighting its reduction in performance compared to the model refined through progressive transfer learning, as detailed in Table \ref{tab:compare_train}. A notable finding from this assessment was the increased reconstruction loss presented by the model trained from scratch, underscoring its limited ability to precisely reconstruct 3D metamaterial units from their latent representations. 
In addition, we compare the computational cost associated with both training methodologies (Table \ref{tab:compare_train}). The progressive transfer learning-based training strategy incurs higher computational demands, with a computational cost 88.4\% greater than that of the model trained from scratch. This increased cost is attributed to the need for multiple runs to fine-tune the loss term coefficients optimally. It is also worth noting that, extending the training epochs for the model trained from scratch (e.g., using the same training epochs as the model trained through progressive transfer learning) does not lead to any improvement in its accuracy.

\begin{table}[H]
  \centering
  \caption{Comparison of the proposed progressive transfer learning-based training and the training from scratch. The reconstruction loss, KL divergence loss, and regression loss for both the training set and validation set are reported.}
  \label{tab:compare_train}
    \begin{tabular}{lcc}
    \toprule
    & \multicolumn{2}{c}{Training Method} \\  
    \cmidrule{2-3}
    & \makecell{Progressive \\ Transfer Learning} & From Scratch \\  
    \midrule
    Reconstruction loss wt. & \multicolumn{2}{c}{1} \\
    \midrule
    KL loss wt. (\(\alpha_2\)) & \multicolumn{2}{c}{$1 \times 10^{-3}$} \\
    \midrule
    Regression loss wt. (\(\alpha_3\)) & \multicolumn{2}{c}{$1 \times 10^{-3}$} \\
    \midrule
    Recon. MSE training Loss & 0.0089 & 0.0265 \\  
    \midrule
    Recon. MSE val. Loss & 0.0105 & 0.0280 \\  
    \midrule
    KL training Loss & 2.686 & 2.570 \\ 
    \midrule
    KL val. Loss & 2.594 & 2.632 \\  
    \midrule
    Reg. NLL training Loss & -3.567 & -3.477 \\  
    \midrule
    Reg. NLL val. Loss & -2.797 & -2.883 \\  
    \midrule
    Computational Cost & \(\sim 442.1\) minutes & \(\sim 234.6\) minutes \\  
    \bottomrule
    \end{tabular}%
\end{table}

\begin{figure}[H]
\centering
\includegraphics[width=0.7\linewidth]{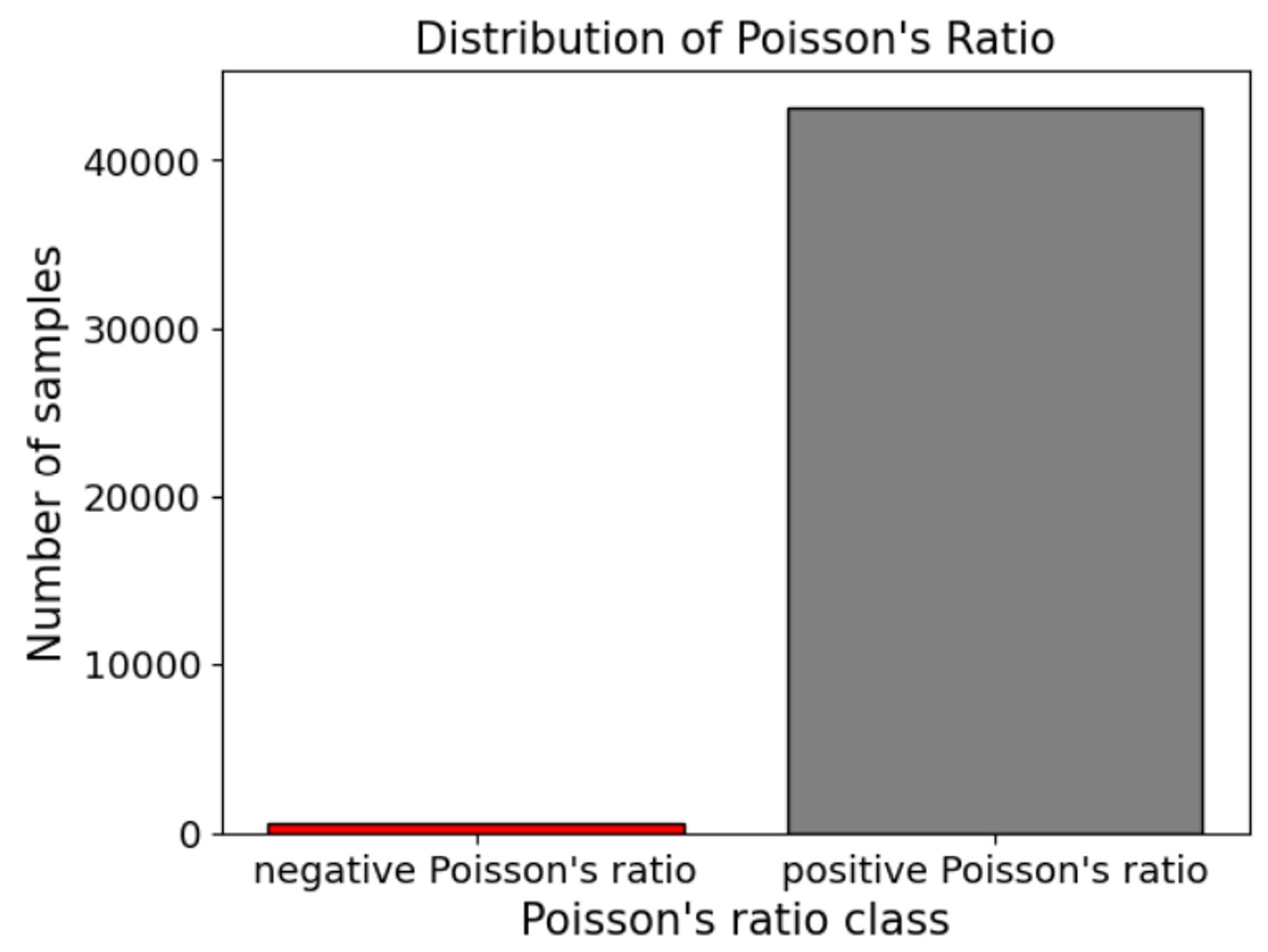}
\caption{\label{fig:pr_distribution}Distribution of negative and positive Poisson's ratio samples in the dataset.}
\end{figure}

It is important to highlight that, in addition to the progressive transfer learning-based training strategy, we implemented a down-selection technique to address data imbalance. As illustrated in Figure \ref{fig:pr_distribution}, the dataset for Poisson's ratio is unbalanced. We retained the original data in the test and validation sets, while down-selecting the data with positive Poisson's ratio in the training set by randomly removing a portion of the data. The data with positive Poisson's ratio in the training set was reduced to 20\%, 30\%, 40\%, 50\%, 60\%, 70\%, and 80\% of its original size, while the data with negative Poisson's ratio was kept unchanged. Ultimately, we selected 60\% as the down-selection portion, as it provided the best validation accuracy during model training.

\subsubsection{Performance assessment}\label{subsubsec:performance}
After successfully training the uncertainty-aware deep learning framework, the model is evaluated in two aspects: the reconstruction accuracy of the VAE model and the prediction accuracy of the regression model.

\begin{figure}[H]
\centering
\includegraphics[width=0.9\linewidth]{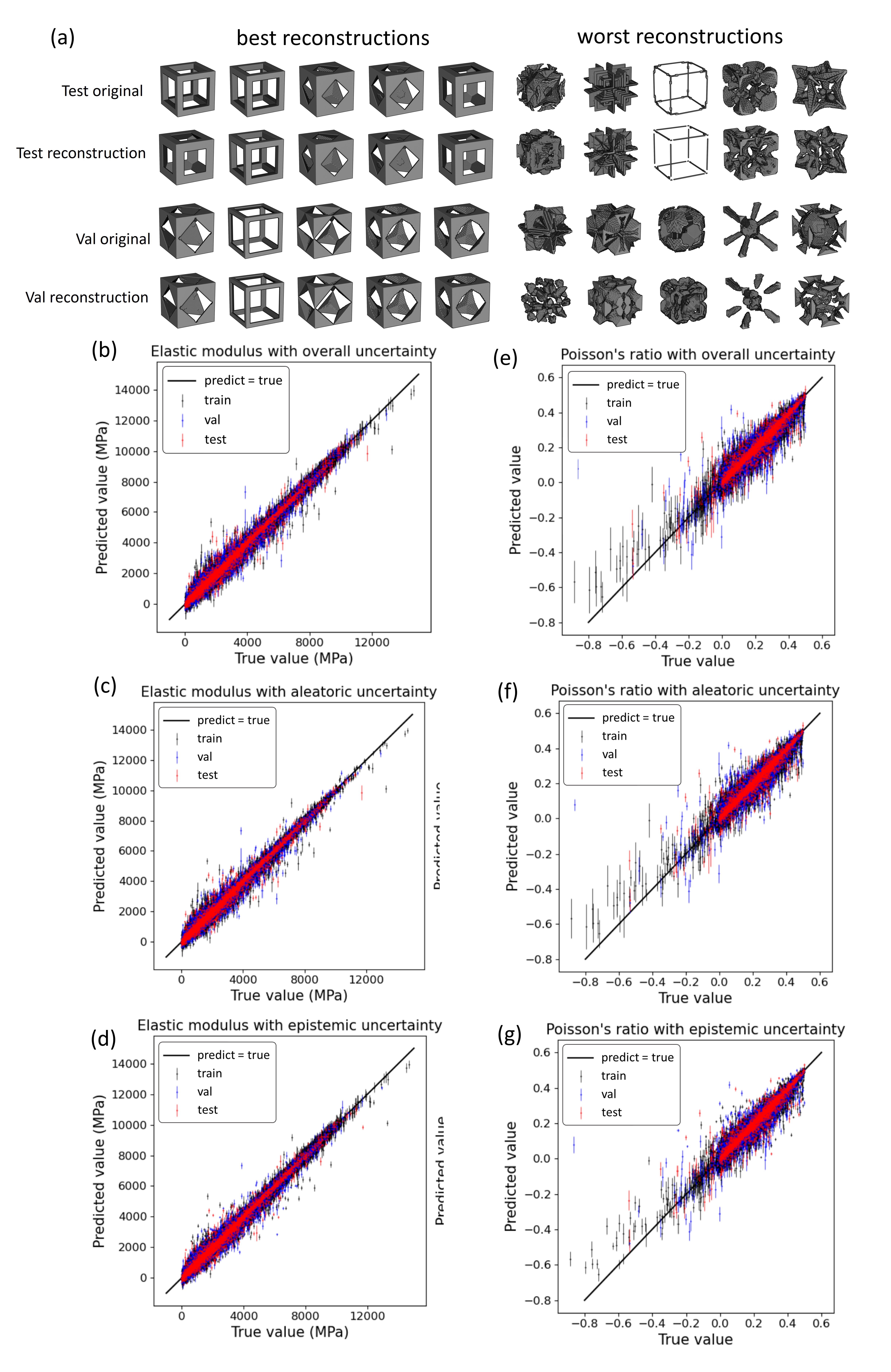}
\caption{\label{fig:validation} (a) Original and reconstructed structures for the five best and five worst cases in the test set and validation set; Comparison of predicted elastic modulus versus the true elastic modulus, with (b) predicted overall uncertainty, (c) predicted aleatoric uncertainty, and (d) predicted epistemic uncertainty; Comparison of predicted Poisson's ratio versus the true Poisson's ratio, with (e) predicted overall uncertainty, (f) predicted aleatoric uncertainty, and (g) predicted epistemic uncertainty.}
\end{figure}

The reconstruction accuracy of VAE is evaluated by a voxelated comparison of the original structure and the reconstructed structure. For a better visual comparison, we showcase the top five best reconstruction cases and top five worst reconstruction cases in the validation set and test set (Figure \ref{fig:validation}a). 
We discover that structures with detailed information, such as shell structures, generally exhibit lower reconstruction accuracy. Conversely, structures characterized by simple geometric features tend to demonstrate higher reconstruction accuracy.
We define the metric of the reconstruction accuracy as the percentage of the correctly predicted voxels over the structure domain:

\begin{equation}\label{eq:voxelerror}
\delta_{\text{recon}} = \left(\frac{1}{N_{sample} \times l^3} \sum_{n=1}^{N_{sample}} \sum_{i=1}^{l} \sum_{j=1}^{l} \sum_{k=1}^{l} \left| \bm{O}_{ijk}^{(n)} - \bm{R}_{ijk}^{(n)} \right| \right) \times 100\%
\end{equation}
where $N_{sample}$ represents the number of structures analyzed, which can be the number of data in training, validation, or test datasets; $l$ represents the voxel length of the structures, with $l=48$ in our dataset.  $\bm{O}_{ijk}^{(n)}$ and $\bm{R}_{ijk}^{(n)}$ represent the original and reconstructed voxel values at position $(i, j, k)$ for the $n$-th structure, respectively.

Following the outlined sampling method mentioned in section \ref{subsec:uncertainty}, we calculate the mean $\mu(f(\bm{z}_i))$ and overall uncertainty $\sigma(f(\bm{z}_i))$ for predicted properties corresponding to each latent vector $\bm{z}_i$ in train, test and validation set, using equations (\ref{eq:cal_mean})-(\ref{eq:cal_total_uncertainty}). Our analysis primarily concentrates on the accuracy of mean value predictions made by the property predictor. This focus is due to the complexity arising from the mixed uncertainties in standard deviation estimates, complicating the separation and measurement of distinct uncertainty factors. The property predictor's accuracy is assessed using the coefficient of determination ($R^2$) and the normalized root mean squared error (NRMSE). The $R^2$ measures how far the observed data deviate from their true mean:

\begin{equation}\label{eq:R2error}
R^2 = 1 - \frac{\sum_{i=1}^{N_{sample}} (\bm{Y}_i - \hat{\bm{Y}}_i)^2}{\sum_{i=1}^{N_{sample}} (\bm{Y}_i - \bar{\bm{Y}}_i)^2}
\end{equation}
while the NRMSE measures the average difference between values predicted by the model and the actual values: 
\begin{equation}\label{eq:NRMSEerror}
\text{NRMSE} = \frac{1}{\max(\bm{Y}) - \min(\bm{Y})} \sqrt{\frac{1}{N_{sample}} \sum_{i=1}^{N_{sample}} (\hat{\bm{Y}}_i - \bm{Y}_i)^2}
\end{equation}
where $\bm{Y}_i$ represents the true response of the $i$-th sample, $\hat{\bm{Y}}_i$ represents the predicted response of the $i$-th sample, $\bar{\bm{Y}}_i$ is the mean value defined by $\bar{\bm{Y}}_i = \frac{1}{N_{\text{sample}}} \sum_{i=1}^{N_{\text{sample}}} \bm{Y}_i$, $\max(\bm{Y})$ represents the maximum value of $\bm{Y}$ in training set and validation set, $\min(\bm{Y})$ represents the minimum value of $\bm{Y}$ in training set and validation set.
A higher $R^2$ value and a lower NRMSE value indicate a more accurate model. It is to be noted that, the true responses are calculated for each data in the datasets with no property variations in finite element simulations. The prediction accuracies of mean values are shown in Table \ref{tab:accuracy}. The predicted overall uncertainty is calculated by Equation \ref{eq:cal_alea}-\ref{eq:cal_total_uncertainty}, as illustrated in Figure \ref{fig:validation}b-\ref{fig:validation}g.

\begin{table}[htbp]
  \centering
  \caption{Reconstruction accuracy of the deep generative model and prediction accuracies of the property predictor.}
    \begin{tabular}{cccccc}
    \toprule
    & \multicolumn{2}{c}{Reconstruction Accuracy} & \multicolumn{3}{c}{Property Prediction} \\
    \cmidrule(lr){2-3} \cmidrule(lr){4-6}
    & Metric & Value & Metric & $E$ & $\nu$ \\
    \midrule
    \multirow{2}{*}{Training Set} & \multirow{2}{*}{$\delta_{\text{recon}}$} & \multirow{2}{*}{0.9833} & $R^2$ & 0.9932 & 0.9795 \\
                                  & & & NRMSE & 0.0114 & 0.0180 \\
    \midrule
    \multirow{2}{*}{Validation Set}  & \multirow{2}{*}{$\delta_{\text{recon}}$} & \multirow{2}{*}{0.9823} & $R^2$ & 0.9862 & 0.9449 \\
                                     & & & NRMSE & 0.0167 & 0.0233 \\
    \midrule
    \multirow{2}{*}{Test Set}  & \multirow{2}{*}{$\delta_{\text{recon}}$} & \multirow{2}{*}{0.9824} & $R^2$ & 0.9857 & 0.9435 \\
                               & & & NRMSE & 0.0171 & 0.0226 \\
    \bottomrule
    \end{tabular}%
  \label{tab:accuracy}%
\end{table}%

\begin{figure}[H]
\centering
\includegraphics[width=1\linewidth]{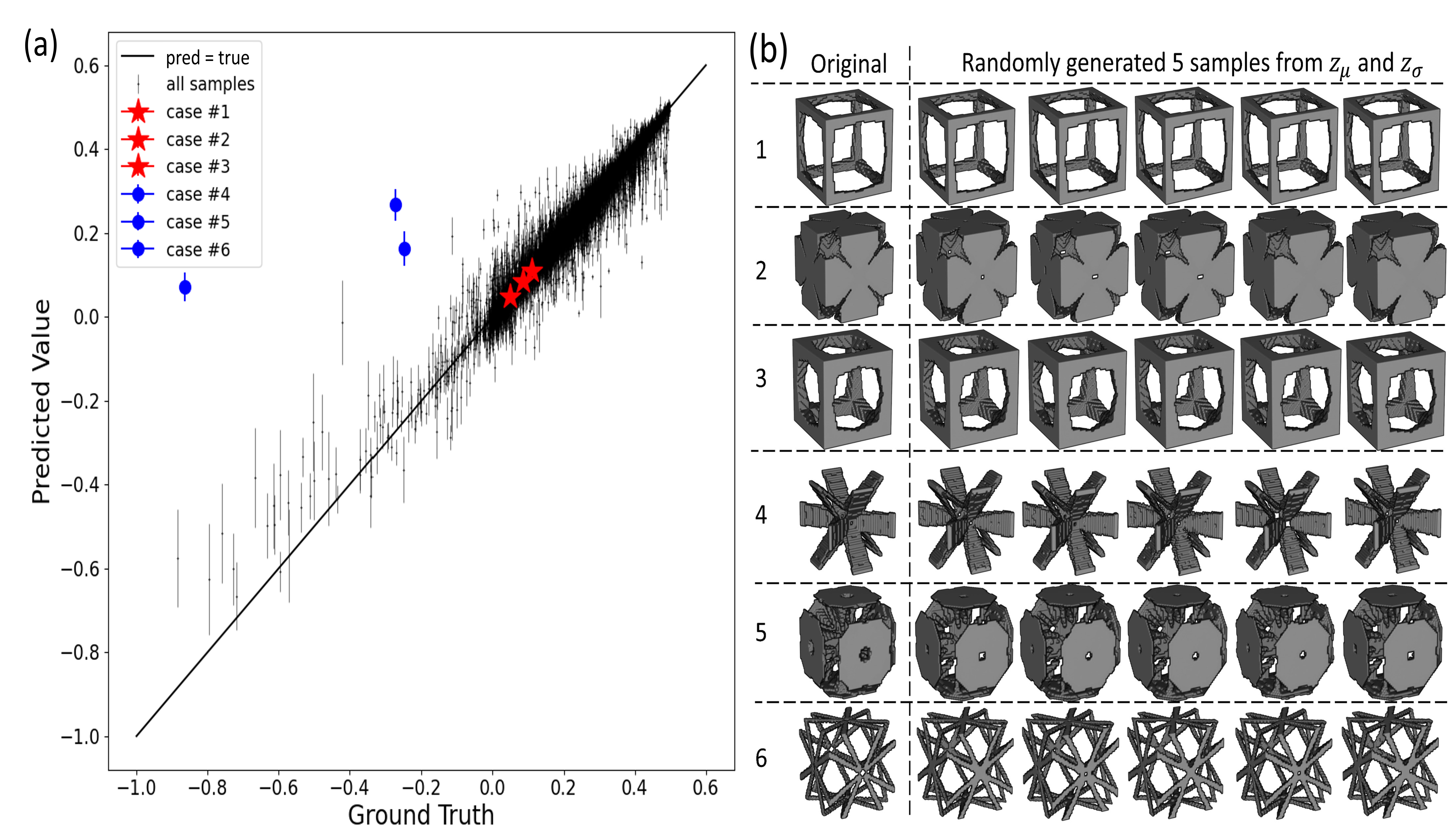}
\caption{\label{fig:shape_variation} (a) Selected three best-predicted cases and three worst-predicted cases. (b) The original metamaterial structure in the original dataset and the reconstructed metamaterial structure by sampling in the latent space through $\bm{z}_{\mu_i}$ and $\bm{z}_{\sigma_i}$. 1-3 are the three best predictions, 4-6 are the three worst predictions.}
\end{figure}

As shown in Figure \ref{fig:validation}b-\ref{fig:validation}g, most of the data in the train, validation, and test set strictly adhere to the 45-degree line (a line that shows equality between the true and predicted values). However, a few data points deviated a lot from the 45-degree line.
To investigate the causes of these poor predictions, we selected the three best (cases 1-3) and three worst (cases 4-6) predicted samples (shown in Figure \ref{fig:shape_variation}a) from all datasets for the $\nu$ prediction, identifying their corresponding latent space values $\bm{z}_{\mu_i}$ and $\bm{z}_{\sigma_i}$. For each of these six cases, we sampled from their $\bm{z}_{\mu_i}$ and $\bm{z}_{\sigma_i}$ and generated 80 different latent vector realizations. These latent vectors were then decoded to the original structure space, resulting in 80 unique geometrical realizations for each case. To illustrate this variation, Figure \ref{fig:shape_variation}b displays five randomly selected structures for each case, all representing the same type of metamaterial unit but with distinct geometrical variations.

We quantify these variations in geometry by the relative voxel difference ($\epsilon_{\text{relative}}$). 
The relative voxel difference $\epsilon_{\text{relative}}$, on the other hand, measures the voxel differences normalized by the magnitude of the original voxel values, thus providing a scale-independent measure of the variation, defined as:

\begin{equation}\label{eq:relative_voxelerror}
\epsilon_{\text{relative}} = \frac{1}{N} \sum_{n=1}^{N} \sum_{i=1}^{l} \sum_{j=1}^{l} \sum_{k=1}^{l} \frac{\left| \bm{O}_{ijk}^{(n)}  - \bm{R}_{ijk}^{(n)} \right|}{|\bm{O}_{ijk}^{(n)} |}
\end{equation}
where $N=80$ represents the total number of sampling points in the latent space for $\bm{z}_{\mu_i}$ and $\bm{z}_{\sigma_i}$. $\bm{O}_{ijk}^{(n)}$ and $\bm{R}_{ijk}^{(n)}$ represent the original and reconstructed voxel values at position $(i, j, k)$ for the $n$-th generated structure, respectively. As detailed in Table \ref{tab:abs_rel_errors}, cases 4-6 exhibit higher relative voxel differences compared to cases 1-3, indicating worse reconstruction accuracy. Poor reconstruction accuracy in these cases would result in higher errors in property predictions and higher predicted uncertainties. We also calculate these samples' corresponding true aleatoric uncertainty by sampling multiple material properties and performing multiple FEA simulations on the same structure. As reported in Table \ref{tab:abs_rel_errors}, the predicted aleatoric uncertainties for cases 1-3 align with their true aleatoric uncertainties, whereas cases 4-6 show significantly higher predicted aleatoric uncertainties compared to their true values. This overestimation is likely due to the errors in the model's function approximation. Since aleatoric uncertainty is defined as data noise, the predicted aleatoric uncertainty is assumed to be influenced solely by the noise in the data. However, in practice, errors can also arise from the model's ability to approximate the true function accurately \parencite{seitzer2022pitfalls,yang2023explainable}. When the model's predictions deviate significantly from the true values due to its limitations in capturing the underlying relationships, these approximation errors contribute to aleatoric uncertainty.

\begin{table}[H]
  \centering
  \caption{Comparison of metrics for the best and worst prediction cases. The metrics include relative voxel difference $\epsilon_{\text{relative}}$, error in Poisson's ratio prediction $\mu(\nu)$, predicted total uncertainty $\sigma_{\text{total}}(\nu)$, predicted aleatoric uncertainty $\sigma_{\text{aleatoric}}(\nu)$, predicted epistemic uncertainty $\sigma_{\text{epistemic}}(\nu)$, and true aleatoric uncertainty $\sigma_{\text{aleatoric}}(\nu)$.}
  \resizebox{\textwidth}{!}{
    \begin{tabular}{ccccccc}
    \toprule
    Sample $\#$ & $\epsilon_{\text{relative}}$ & \makecell{Error in \\ $\mu(\nu)$ }& \makecell{Predicted \\ $\sigma_{\text{total}}(\nu)$}  & \makecell{Predicted \\ $\sigma_{\text{epistemic}}(\nu)$} & \makecell{True \\ $\sigma_{\text{aleatoric}}(\nu)$} & \makecell{Predicted \\ $\sigma_{\text{aleatoric}}(\nu)$ }\\
    \midrule
    1 & 0.0176 & 1e-8 & 0.0010 & 0.0008 & 0.0003 & 0.0005 \\
    2 & 0.0306 & 7e-8 & 0.0030 & 0.0020 & 0.0008 & 0.0022 \\
    3 & 0.0081 & 2e-7 & 0.0009 & 0.0005 & 0.0006 & 0.0007 \\
    4 & 0.1856 & 0.4669 & 0.0118 & 0.0103 & 0.0001 & 0.0056 \\
    5 & 0.1603 & 0.2697 & 0.0298 & 0.0206 & 0.0003 & 0.0215 \\
    6 & 0.3175 & 0.2059 & 0.0204 & 0.0053 & 0.0003 & 0.0196 \\
    \bottomrule
    \end{tabular}
  }
  \label{tab:abs_rel_errors}
\end{table}

\subsubsection{Generation of New Metamaterial Units by Sampling on the Latent Feature Space}

\begin{figure}[H]
\centering
\includegraphics[width=0.9\linewidth]{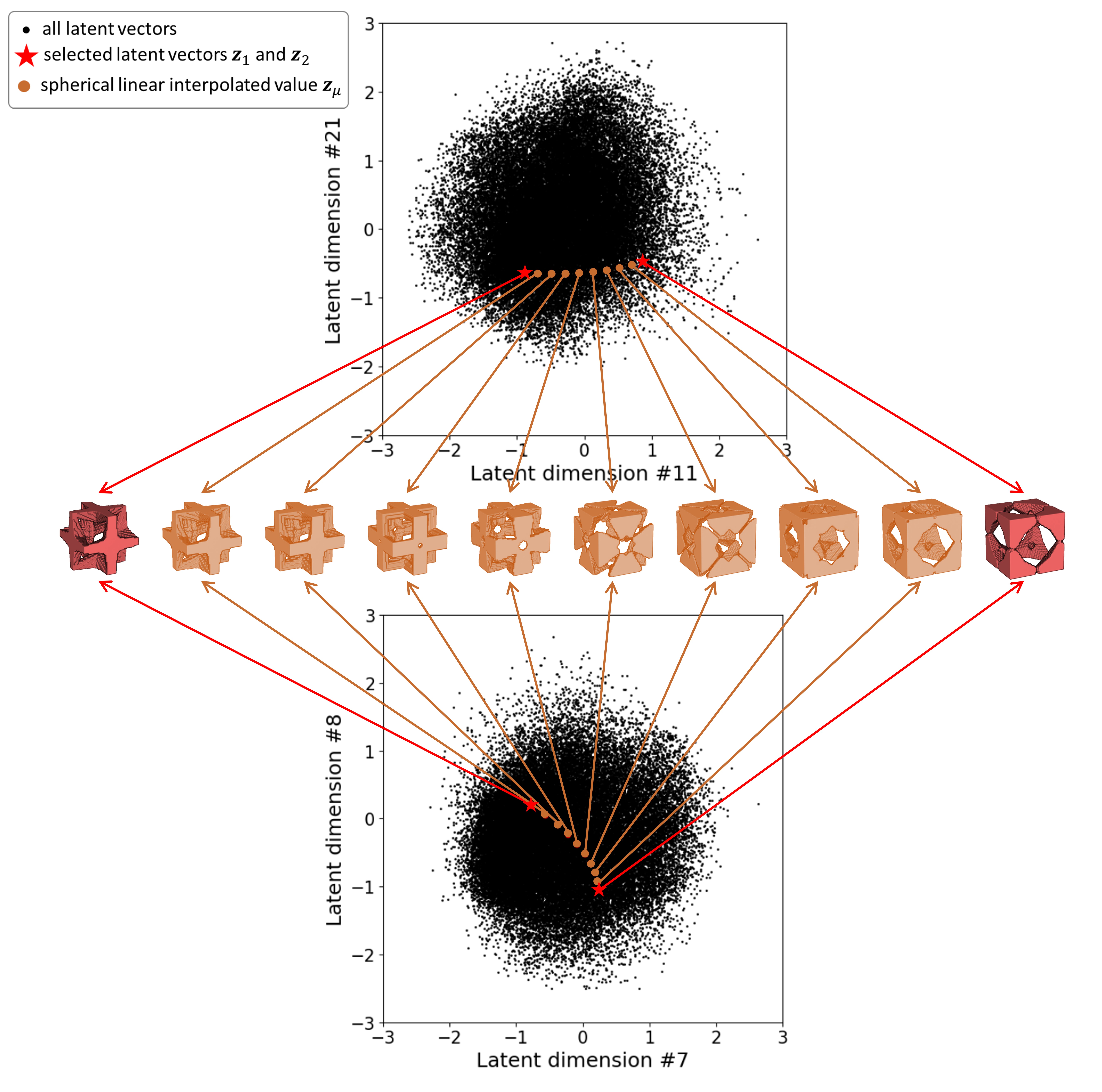}
\caption{\label{fig:interpolation} 
An example of creating evolving metamaterial units by adjusting latent vector values in the latent space using spherical linear interpolation.}
\end{figure}

We illustrate the mechanisms of generating continuously evolving metamaterial unit designs by manipulating the values of latent vectors in the latent feature space. Spherical linear interpolation (slerp), first introduced and applied in \parencite{white2016sampling}, is utilized to interpolate between two points within the latent space. Traditionally, linear interpolation has been favored for its simplicity. However, in the context of a high-dimensional latent space with a Gaussian prior, linear interpolation can result in blurry shapes due to deviations from the model's prior distribution. 
Spherical linear interpolation addresses this issue by ensuring interpolated points are uniformly distributed on the hypersphere and stay within regions consistent with the model's prior distribution, thereby generating more coherent and realistic shapes. The formulation for spherical linear interpolation is given by:

\begin{equation}
\bm{z}_{\mu} = \text{slerp}(\bm{z}_1, \bm{z}_2; \mu) = \frac{\sin{(1-\mu)\theta}}{\sin{\theta}} \bm{z}_1 + \frac{\sin{\mu\theta}}{\sin{\theta}} \bm{z}_2
\end{equation}
where \text{slerp} denotes the spherical linear interpolation operation; $\bm{z}_1$ and $\bm{z}_2$ are two randomly selected latent vectors in the latent feature space; $\mu$ represents the location along the path, with $\mu=0$ indicating the start and $\mu=1$ the end point. $\bm{\theta} = \arccos\left(\frac{\bm{z}_1^T \bm{z}_2}{\|\bm{z}_1\| \|\bm{z}_2\|}\right)$, and $\bm{z}$ follows a normal distribution. Figure \ref{fig:interpolation} demonstrates an example of using spherical interpolation of latent vector values in the latent space to generate metamaterial units. We randomly selected two metamaterial units from our dataset and encoded them to obtain the corresponding latent vectors $\bm{z}_1$ and $\bm{z}_2$. The values $\bm{z}_{\mu}$ are spherical linear interpolated points, which are then decoded to generate continuous metamaterial units not present in the original dataset. Out of the total 32 dimensions in our latent space, four dimensions—latent dimensions \#7, \#8, \#11, and \#21—are randomly selected and grouped in pairs to better visualize the interpolation path.

\subsection{Robust Design of Metamaterial Units}\label{sec5}

In this section, we implement two robust design cases for designing metamaterial units with different objectives: a) bulk modulus maximization and b) elastic modulus and Poisson's ratio maximization. We compared our designs with the designs found by the topology optimization (TO) method in literature and the designs found by a deterministic design optimization that does not account for uncertainty.

\subsubsection{Case 1: Maximization of Bulk Modulus}
In this study, we aim to maximize the bulk modulus of the metamaterial units while simultaneously minimizing the design's uncertainty, targeting a volume fraction in the range of 0.299 to 0.301. 
The bulk modulus ($K$) of a metamaterial unit is given by:
\begin{equation}
K = \frac{E}{3(1-2\nu)}
\end{equation}
where $E$ and $\nu$ are the Elastic Modulus and Poisson's ratio. The robust design optimization problem is stated as follows:
\begin{equation}
\begin{aligned}
&\max_{\bm{z}}  &\mu(K(\bm{z})) - \beta \sigma(K(\bm{z})) \\
&\text{s.t. } &|V_f(\bm{z})-0.3|=0.001 \\
& &\text{min}(\bm{z}) \leq \bm{z} \leq \text{max}(\bm{z})
\end{aligned}
\end{equation}
where $\beta$  represents a weighting factor that adjusts the significance of the mean relative to the standard deviation and $V_f$ is the volume fraction. Elevating the value of $\beta$ enhances the emphasis on reducing variability; when $\beta=0$, the objective function simplifies to determining the lowest expected value for the bulk modulus. Our goal is thus to identify the optimal $\beta$ value that strikes a balance between achieving the desired objective function and managing the total uncertainty as predicted by the deep learning framework.

\begin{figure}[H]
\centering
\includegraphics[width=1\linewidth]{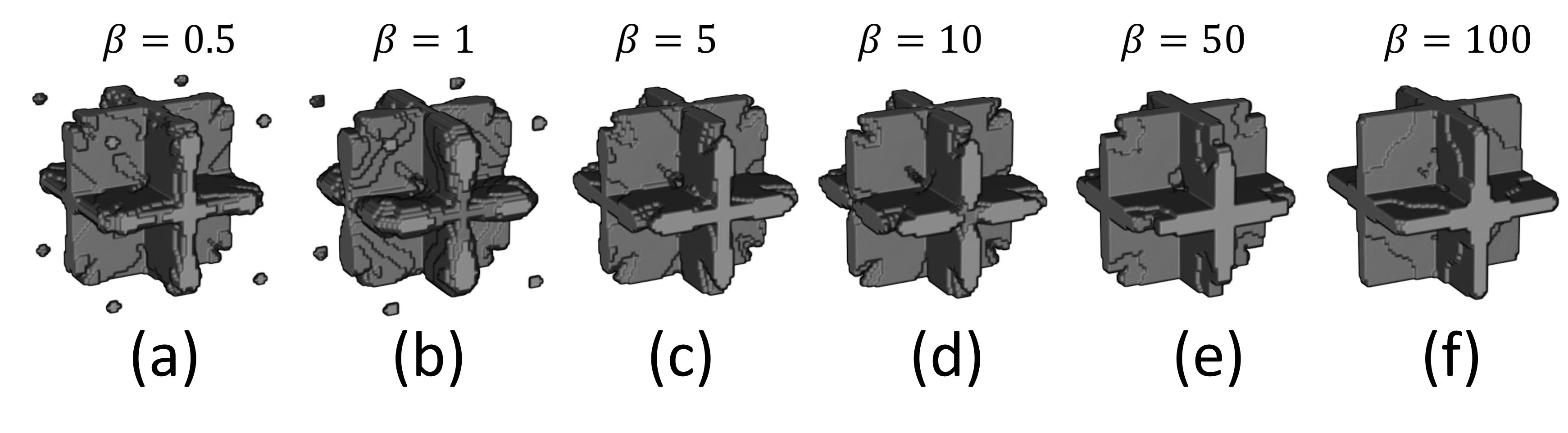}
\caption{\label{fig:design1} Metamaterial units obtained by robust design approach with (a) $\beta=0.5$; (b) $\beta=1$; (c) $\beta=5$; (d) $\beta=10$; (e) $\beta=50$; (f) $\beta=100$.}
\end{figure}

Utilizing the NSGA-II algorithm, we identify the optimal design encoded as a latent vector $\bm{z}$, which is subsequently decoded into a 3D voxel representation of the metamaterial unit. We investigated various $\beta$ values, from 0.5 to 100, and recorded the resulting optimal metamaterial units (Figure \ref{fig:design1}) and their associated uncertainties obtained from the uncertainty-aware deep learning framework. As depicted in Figure \ref{fig:design1}, increasing $\beta$ leads to designs with simpler geometric features and fewer intricate details. The results of each optimization, detailed in Table \ref{tab:addlabel2}, show that both the predicted bulk modulus and the uncertainty decrease with higher $\beta$ values. Notably, structures become more integrated as $\beta$ reaches or exceeds 5. Thus, we chose $\beta = 5$ as the optimal level of uncertainty for inclusion in our robust design approach. We also calculated the optimal design's true bulk modulus by performing FEA simulation. It is to be noted that there exists a large discrepancy between the predicted and true values of the bulk modulus of optimal design with $\beta = 0.5$ and $\beta = 1$. This is due to the optimal metamaterial structure obtained having some floating noise, which would influence the FEA simulation result. Additionally, the predicted aleatoric uncertainty is slightly higher than the true aleatoric uncertainty, particularly in designs with poor bulk modulus predictions. This overestimation is due to errors in the model's function approximation, as discussed in section \ref{subsubsec:performance}.

The optimal design found by our approach is compared with the metamaterial unit design obtained by the method proposed in the literature \parencite{huang2011topological}, which introduces a TO approach for creating metamaterial units with maximized bulk modulus. We select an optimal design at a volume fraction of 0.3 in the literature. For an appropriate and fair comparison, we resize our selected design to the same \(26 \times 26 \times 26\) cubic domain as defined in the literature, and use the same 8-node brick elements in the FEA simulation. As a result, our proposed deep learning framework-based robust design optimization successfully yielded designs with a higher normalized bulk modulus compared to those reported in \parencite{huang2011topological}. Based on the results, we summarize below some strengths as well as limitations of the proposed uncertainty-aware deep learning framework-based robust design optimization relative to TO:
\begin{itemize}
\item After the initial training of the uncertainty-aware deep learning framework, obtaining new designs and their corresponding uncertainties is rapid in the inference stage. In contrast, TO requires significant computational resources due to the iterative process.
\item Explicit consideration of uncertainties is necessary in the design formulation for TO. In contrast, our approach implicitly learns these uncertainties.
\item TO typically focuses on optimizing within predefined parameters and constraints, which might limit the exploration of novel design spaces. In comparison, our approach can explore broader design space and generate novel design configurations by learning complex patterns and relationships from the training data.
\item There are inevitable errors in the predicted property values using our proposed robust optimization method, whereas the property values predicted by TO are accurate.
\end{itemize}

\begin{table}[htbp]
  \centering
  \caption{Comparison of the true and predicted bulk modulus, the associated true and predicted uncertainties, and the volume fraction of the 3D metamaterial optimal design candidate for different values of $\beta$ used in the proposed robust design framework.}
  \resizebox{\textwidth}{!}{%
    \begin{tabular}{cccccccc}%
    \toprule
    & \multicolumn{6}{c}{Objective} & \multicolumn{1}{c}{Constraints} \\
    \cmidrule(lr){2-7} \cmidrule(lr){8-8}
    $\beta$ value & \makecell{True\\ $\mu(K)$ }& \makecell{Predicted\\ $\mu(K)$ }& \makecell{Predicted \\$\sigma_{\text{total}}(K)$ }& \makecell{Predicted\\ $\sigma_{\text{epistemic}}(K)$ }& \makecell{True \\$\sigma_{\text{aleatoric}}(K)$ }& \makecell{Predicted \\$\sigma_{\text{aleatoric}}(K)$ }& $V_f$ \\
    \midrule
    0.5   & 6347.29 & 7422.34 & 449.13 & 371.30 & 61.75 & 252.69 & 0.2996 \\
    1     & 4554.07 & 7321.25 & 379.76 & 345.63 & 45.10 & 157.33 & 0.2992 \\
    \textbf{5}  & \textbf{6300.38} & \textbf{6677.15} & \textbf{370.39} & \textbf{361.60} & \textbf{60.22} & \textbf{80.21} & \textbf{0.2998} \\
    10    & 6299.71 & 6559.23 & 345.51 & 339.68 & 60.21 & 63.18 & 0.2991 \\%
    50    & 5677.37 & 6001.22 & 251.73 & 241.60 & 54.18 & 70.68 & 0.2993 \\
    100   & 5702.07 & 5732.55 & 223.59 & 217.89 & 54.25 & 50.15 & 0.2999 \\
    \bottomrule
    \end{tabular}%
  }
  \label{tab:addlabel2}%
\end{table}

\begin{table}[H]
  \centering
  \caption{Comparison of FEA simulated Bulk Modulus between TO structure and the robust design approach.}
  \begin{tabular}{C{3cm}C{4cm}C{4cm}C{3cm}}
    \toprule
    Design & Optimal Structure & \makecell{FEA simulated \\Bulk Modulus $K$ (MPa)} & Volume Fraction $V_f$ \\
    \midrule
    TO~\parencite{huang2011topological} & \includegraphics[width=0.1\textwidth]{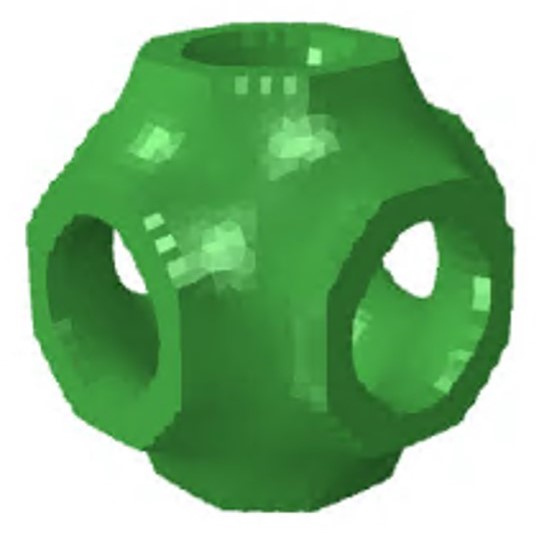} & 5577.8 & 0.3 \\
    Robust & \includegraphics[width=0.1\textwidth]{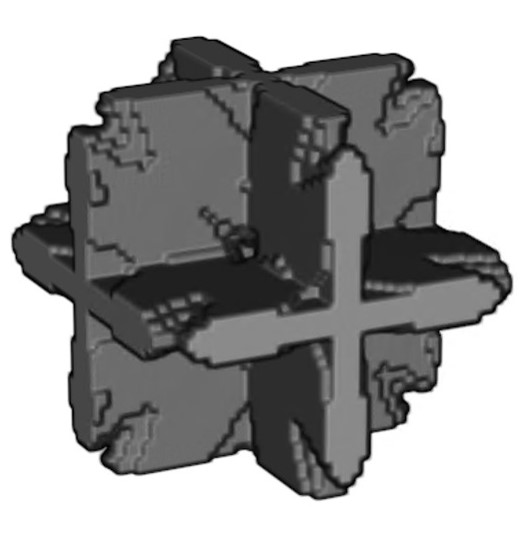} & 5954.8 & 0.302 \\
    \bottomrule
  \end{tabular}%
  \label{tab:fea_bulk_modulus}
\end{table}

\subsubsection{Case 2: Maximization of Elastic Modulus and Poisson's Ratio}\label{subsec:design2}

\begin{figure}[H]
\centering
\includegraphics[width=1\linewidth]{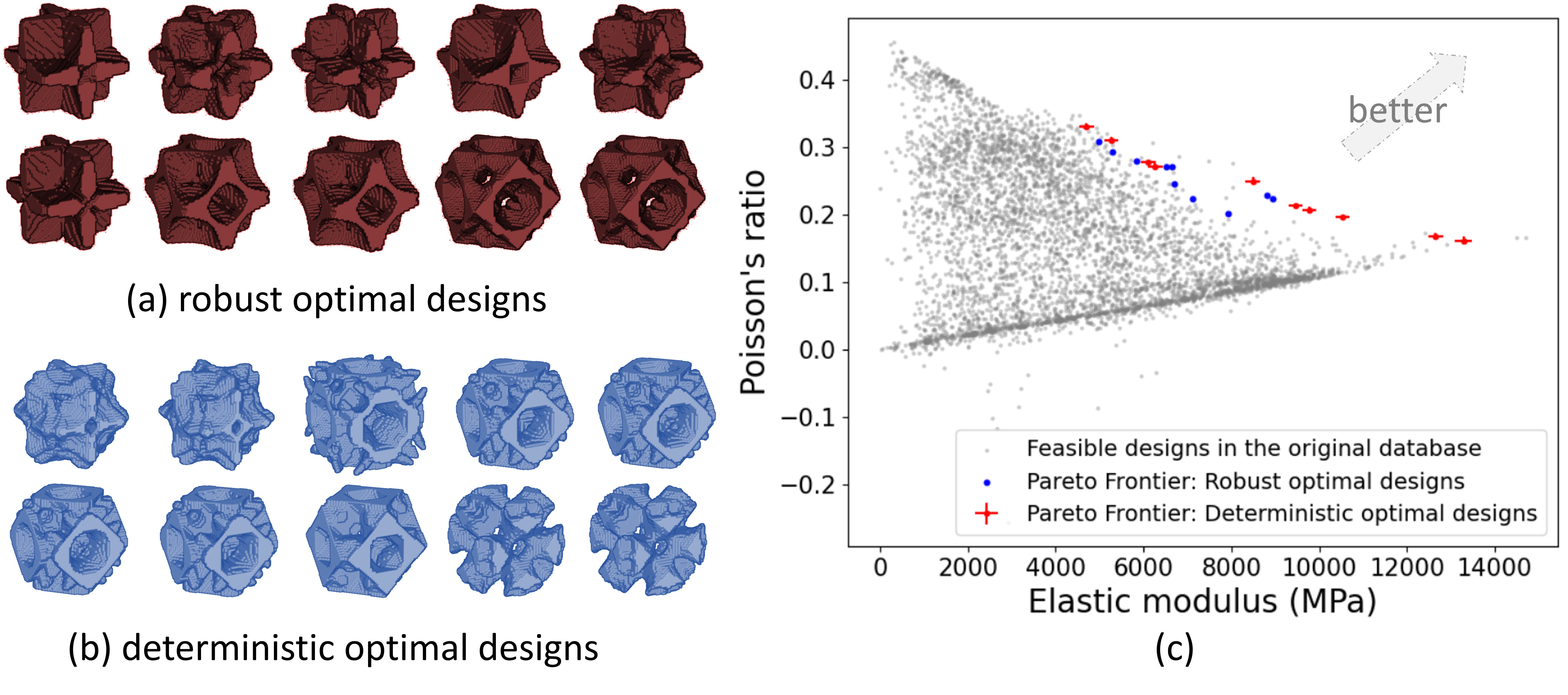}
\caption{\label{fig:design2} Optimal designs obtained from (a) robust design optimization (b) deterministic design optimization. (c) Non-dominated design sets obtained by multi-objective optimization robust design optimization and deterministic design optimization.}
\end{figure}

In this case, a multi-objective robust design optimization of metamaterial units is conducted to maximize the elastic modulus $E$ and Poisson’s ratio $\nu$ simultaneously, with a volume fraction of $0.32$ and considering the associated uncertainty using the proposed approach. From the previous case study, we select $\beta = 5$ as the optimal level of uncertainty for inclusion in our robust design approach. Then the design problem is formulated as follows:

\begin{equation}\label{eq:obj}
\begin{aligned}
\max_{\bm{z}} & \; \;  \{ \mu(E(\bm{z})) - 5\sigma(E(\bm{z})), \; \mu(\nu(\bm{z})) - 5\sigma(\nu(\bm{z})) \} \\
\text{s.t.} & \; \; |V_f(\bm{z}) - 0.32| = 0.001 \\
& \; \; \text{min}(\bm{z}) \leq \bm{z} \leq \text{max}(\bm{z})
\end{aligned}
\end{equation}
NSGA-II is applied to search for the optimal designs (on the Pareto frontier) represented in the form of a latent vector $\bm{z}$. Subsequently, the optimal latent vector is decoded to obtain the metamaterial unit in the format of a 3D voxel image. The obtained optimal metamaterial unit candidates are shown in Figure \ref{fig:design2}a. The true properties of the found designs are verified by simulations. The predicted values and the corresponding ground truth values are compared in Table \ref{table:optimization2}.

The robust design optimization is compared with a deterministic design optimization, where only the mean value of the metamaterial is considered in the design objective formulation. The deterministic design optimization is established based on a deterministic deep learning framework, which follows the same architecture as we proposed in Figure \ref{fig:framework}, but with a feed-forward deep neural network as a property predictor. The feed-forward deep neural network can only capture the mean value of the prediction. None of the uncertainty resources in the deterministic design is considered. The detailed information of the deterministic deep learning framework and its corresponding design optimization is shown in Appendix \ref{sec:appendix1}. The formulation of the deterministic design optimization is expressed as:

\begin{equation}
\begin{aligned}
\max_{\bm{z}} & \;\; \{ E(\bm{z}), \nu(\bm{z}) \} \\
\text{s.t.} & \; \;|V_f(\bm{z}) - 0.32| = 0.001 \\
& \; \; \text{min}(\bm{z}) \leq \bm{z} \leq \text{max}(\bm{z})
\end{aligned}
\end{equation}
The NSGA-II algorithm is employed to identify the optimal latent features, which are subsequently decoded into the optimal 3D voxelated metamaterial units. As depicted in Figure \ref{fig:design2}b, the structures derived from deterministic design optimization exhibit more intricate characteristics and a greater number of small features. Due to the nature of VAEs, which often generate images with blurred borders \parencite{dai2019diagnosing}, these detailed features may not be accurately generated, potentially leading to reduced reliability of the final optimal designs. This is evidenced by the greater discrepancies between the predicted and true properties in the deterministic design candidates (Table \ref{table:optimization2}).

\begin{figure}[H]
\centering
\includegraphics[width=0.9\linewidth]{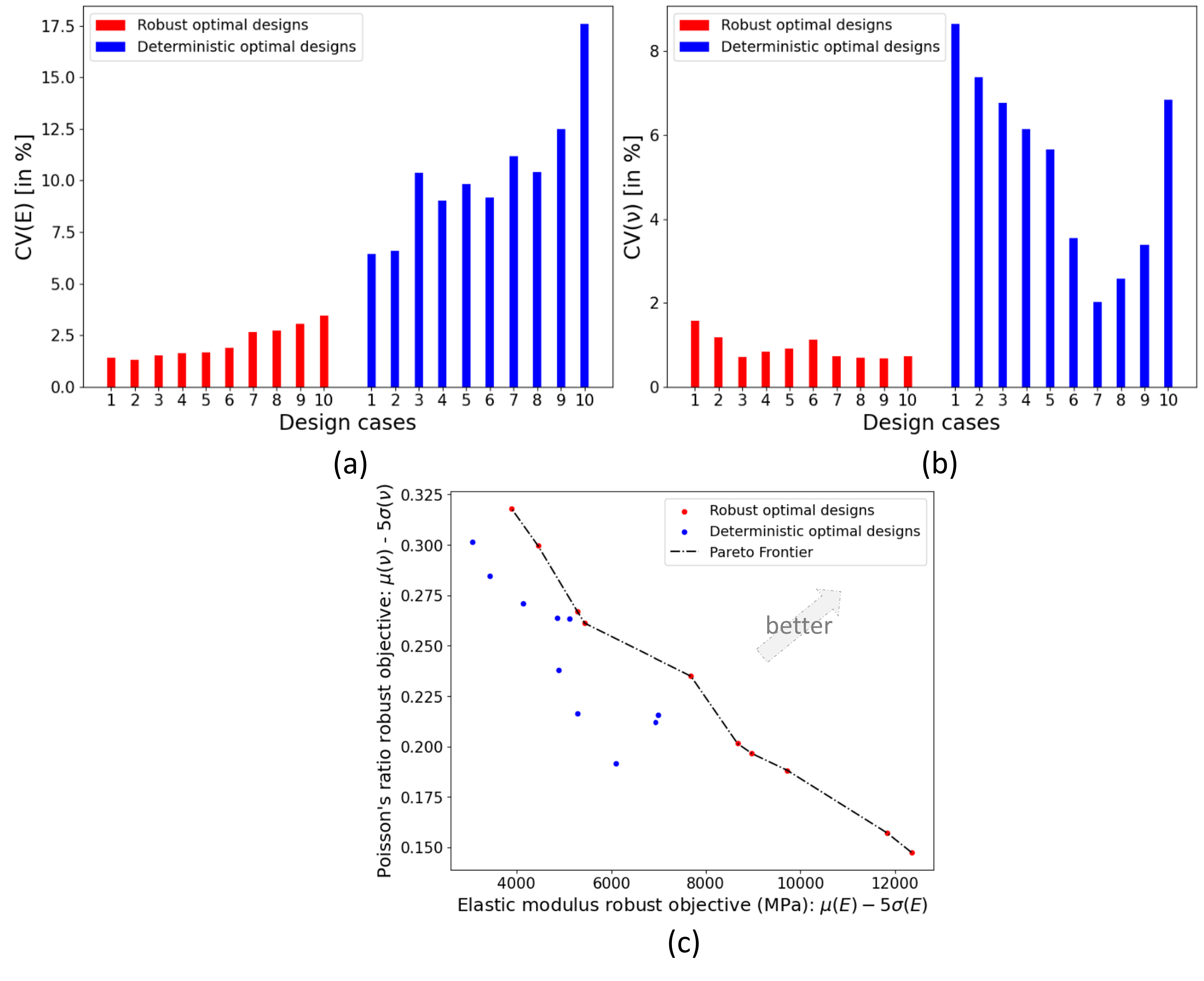}
\caption{\label{fig:CV} (a) Comparison of Elastic modulus CV values between the robust optimal design cases and the deterministic optimal design cases; (b) Comparison of Poisson's ratio CV values between the robust optimal design cases and the deterministic optimal design cases; (c) Robust objective values of the robust and deterministic optimal designs, evaluated using the uncertainty-aware deep learning framework.}
\end{figure}

The true Pareto Frontiers derived from both robust and deterministic design optimizations are presented in Figure \ref{fig:design2}c. 
The discrepancy in the Pareto Frontiers can be attributed to inherent differences in how the optimization methods account for uncertainties. Robust optimization, designed to minimize the impact of uncertainties while maximizing the design objective, yields solutions in regions of the design space with lower uncertainties. In contrast, deterministic optimization, which does not account for uncertainties, results in design candidates with no assurance of low uncertainty.
To validate this observation, we evaluated the relative variability of the design optimization results using the coefficient of variation (CV), defined as the ratio of the standard deviation to the mean, expressed as a percentage. As shown in Figure \ref{fig:CV}a and Figure \ref{fig:CV}b, the CV for both the elastic modulus and Poisson's ratio of the robust design candidates is smaller than that of the deterministic design candidates, indicating that the robust design optimization produces more consistent and reliable outcomes. 
Subsequently, we assessed the robust objective values (Equation \ref{eq:obj}) of these deterministic design candidates, as shown in Figure \ref{fig:CV}c. Compared to the robust objective values of the robust design candidates, the deterministic design candidates fall within the region of dominated sets. Consequently, these designs will not be selected as points on the Pareto Frontier.

In conclusion, our proposed uncertainty-aware deep learning framework-based robust design optimization offers several advantages over deterministic design optimization:

\begin{itemize}
    \item Deterministic optimization produces intricate features that VAEs often struggle to capture accurately, reducing design reliability. Robust optimization ensures that features are well-represented and reliable.
    \item Robust design optimization targets regions with lower uncertainties in the design space, resulting in more reliable designs. In contrast, deterministic design optimization leads to designs with higher uncertainty.
\end{itemize}

\begin{table}[H]
  \centering
  \caption{Comparison of true and predicted material properties of 3D metamaterial optimal design candidates using the proposed robust design approch and the determinsitic design approach}
  \label{table:optimization2}
  \resizebox{\textwidth}{!}{%
    \begin{tabular}{ccccccccc}
    \toprule
    \multicolumn{9}{c}{Proposed Robust Design Optimization} \\
    \midrule
    \multirow{2}[4]{*}{} & \multicolumn{6}{c}{Objective}                 & \multicolumn{2}{c}{Constraints} \\
\cmidrule{2-9}          & \multicolumn{3}{c}{Elastic Modulus $E$ (MPa)} & \multicolumn{3}{c}{Poisson's ratio $\nu$} &  \multicolumn{2}{c}{Volume Fraction } \\
    \midrule
    Case      &\makecell{ Predicted \\$\mu(E)$} & \makecell{True \\$\mu(E)$} & \makecell{Predicted \\$\sigma(E)$ }& \makecell{Predicted \\$\mu(\nu)$} & \makecell{True \\$\mu(\nu)$} & \makecell{Predicted \\$\sigma(\nu)$} & \multicolumn{2}{c}{True} \\
    \midrule
    1     & 12448.3 & 13280.9 & 187.25 & 0.1831 & 0.1709 & 0.0027 & \multicolumn{2}{c}{0.3209} \\
    \midrule
    2     & 12015.5 & 12649.1 & 164.19 & 0.1903 & 0.1776 & 0.0021 & \multicolumn{2}{c}{0.3209} \\
    \midrule
    3     & 9969.8 & 10526.5 & 160.21 & 0.2321 & 0.2261 & 0.0016 & \multicolumn{2}{c}{0.3208} \\
    \midrule
    4     & 9632.5 & 9775.6 & 159.96 & 0.2445 & 0.2365 & 0.0020 & \multicolumn{2}{c}{0.3207} \\
    \midrule
    5     & 9421.8 & 9461.3 & 157.75 & 0.2502 & 0.2425 & 0.0022 & \multicolumn{2}{c}{0.3207} \\
     \midrule
     6     & 8396.5 & 8487.1 & 159.29 & 0.2508 & 0.2488 & 0.0028 & \multicolumn{2}{c}{0.3206} \\
     \midrule
     7     & 6332.0 & 6267.3 & 165.93 & 0.2901 & 0.2711 & 0.0020 & \multicolumn{2}{c}{0.3206} \\
     \midrule
     8     & 6028.4 & 6111.1  & 166.22 & 0.2912  & 0.2765 & 0.0019 & \multicolumn{2}{c}{0.3206} \\
     \midrule
     9     & 5322.7 & 5258.6 & 160.69 & 0.3059 & 0.3105 & 0.0021 & \multicolumn{2}{c}{0.3206} \\
     \midrule
     10    & 4909.1 & 4704.8 & 162.38 & 0.3486 & 0.3320 & 0.0024 & \multicolumn{2}{c}{0.3206} \\
    \midrule
    \midrule
    \multicolumn{9}{c}{Deterministic Design Optimization} \\
    \midrule
    \multirow{2}[4]{*}{} & \multicolumn{6}{c}{Objective}                 & \multicolumn{2}{c}{Constraints} \\
\cmidrule{2-9}          & \multicolumn{3}{c}{Elastic Modulus $E$ (MPa)} & \multicolumn{3}{c}{Poisson's ratio $\nu$} &  \multicolumn{2}{c}{Volume Fraction } \\
    \midrule
    Case      & \makecell{Predicted\\ $\mu(E)$} & \makecell{True \\$\mu(E)$} &\makecell{ Predicted \\$\sigma(E)$} & \makecell{Predicted\\ $\mu(\nu)$  } & \makecell{True \\$\mu(\nu)$ }& \makecell{Predicted \\$\sigma(\nu)$}  & \multicolumn{2}{c}{True} \\
    \midrule
    1     & 9032.8 & 8734.8 &  -    & 0.2287 & 0.2023 & -     & \multicolumn{2}{c}{0.3201} \\
    \midrule
    2     & 8977.3 & 8718.7 &   -   & 0.2333 & 0.1976 & -     & \multicolumn{2}{c}{0.3198} \\
    \midrule
    3    & 8433.2 & 7914.0  &  -    & 0.2402 & 0.1712 & -    &  \multicolumn{2}{c}{0.3196} \\
    \midrule
    4    & 8644.2 & 7112.3 & -     & 0.2442 & 0.2037 & -     & 
    \multicolumn{2}{c}{0.3206} \\
    \midrule
    5     & 6721.8 & 6651.1 & -     & 0.2444 & 0.2508 & -     & \multicolumn{2}{c}{0.3209} \\
     \midrule
     6     & 6655.4 & 6285.5 & -     & 0.2611 & 0.2517 & -     & \multicolumn{2}{c}{0.3208} \\
     \midrule
    7     & 6635.4 & 5843.9 & -     & 0.2632 & 0.2581 & -     & \multicolumn{2}{c}{0.3205} \\
    \midrule
    8    & 6533.2 & 6694.2 & -     & 0.2674 & 0.2453 & -     &  \multicolumn{2}{c}{0.3210} \\
    \midrule
    9     & 6317.5 & 6512.2 & -     & 0.2732 & 0.2512 & -     & \multicolumn{2}{c}{0.3208} \\
     \midrule   
    10     & 5635.8 & 4968.1 & -     & 0.3022 & 0.2776 & -    & \multicolumn{2}{c}{0.3197} \\     
    \bottomrule
    \end{tabular}%
    }
\end{table}

\section{Conclusion}\label{sec:conclusion}

In this study, we introduce a robust design approach using an uncertainty-aware deep learning framework for creating optimal metamaterial units. Both aleatoric and epistemic sources of uncertainties are characterized within the deep learning framework. The proposed approach enables the robust design of metamaterial units by maximizing the mean value of the property and minimizing its associated uncertainty. Our key findings are as follows:

(1) Our uncertainty-aware deep learning framework successfully measures data uncertainty and latent space uncertainty by generating different realizations on the latent feature space.

(2) We demonstrate that our proposed progressive transfer learning-based training strategy is effective in optimizing the weight coefficients of different loss terms as well as the network weights in the uncertainty-aware deep learning framework.

(3) The proposed uncertainty-aware deep learning framework-based design optimization is effective in the robust design of metamaterial units. The efficacy of the proposed design approach is validated by two design cases.

We also identify the limitations of this work:

(1) In this work, we selected the MDN network as the PDNN model. However, alternative PDNN models such as Bayesian neural networks, Monte Carlo Dropout based networks, among others, could also be integrated into the framework. As part of a future work, we aim to incorporate and compare the performance of various types of PDNN models within the proposed deep learning framework.

(2) Unavoidable discrepancies persist between predicted and true responses in optimal designs. These discrepancies may arise from various sources, including data quality, model architecture, and the inherent stochasticity of optimization algorithms in deep neural networks. Consequently, complete elimination of these discrepancies remains unattainable.

\appendix

\renewcommand\theequation{\Alph{section}\arabic{equation}} 
\counterwithin*{equation}{section} 
\renewcommand\thefigure{\Alph{section}\arabic{figure}} 
\counterwithin*{figure}{section} 
\renewcommand\thetable{\Alph{section}\arabic{table}} 
\counterwithin*{table}{section} 

\clearpage
{
\noindent
\LARGE\textbf{Appendix}
}

\begin{appendices}
\section{Hyperparameters of the deep learning model}\label{sec:appendix2}

\begin{table}[H]
  \centering
  \caption{The detailed structure of the encoder, decoder, MDN regressor of the proposed uncertainty-aware deep learning model, and the DNN regressor of the deterministic deep learning model. }
    \begin{tabular}{ll}
    \toprule
    \multicolumn{2}{l}{\textbf{Encoder}} \\
    \midrule
    Block & Specifications \\
    \midrule
    Encoder Conv3d-1 & (Conv32 + ReLU) ×3 + MaxPooling \\
    \midrule
    Encoder Conv3d-2 & (Conv64 + ReLU) ×3 + MaxPooling \\
    \midrule
    Encoder Conv3d-3 & (Conv96 + ReLU) ×3 + MaxPooling \\
    \midrule
    Encoder FC & 2592 + ReLU -> 1000 + ReLU -> 100 \\
    \midrule
    Mean, Variance, Latent vector & 32 \\
    \midrule
    \multicolumn{2}{l}{\textbf{Decoder}} \\
    \midrule
    Block & Specifications \\
    \midrule
    Decoder FC & 32 + ReLU -> 1000 + ReLU -> 2592 \\
    \midrule
    Decoder ConvTranspose3d-1 & (Conv96 + ReLU) × 3 + Upsampling \\
    \midrule
    Decoder ConvTranspose3d-2 & (Conv64 + ReLU) × 3 + Upsampling \\
    \midrule
    Decoder ConvTranspose3d-3 & (Conv32 + ReLU) × 3 + Upsampling \\
    \midrule
    Decoder ConvTranspose3d-4 & (Conv16 + ReLU) × 2 + Conv16 + Sigmoid \\
    \midrule
    \multicolumn{2}{l}{\textbf{MDN Property Predictor}} \\
    \midrule
    Block & Specifications \\
    \midrule
    Property Predictor FC & 256 + ReLU -> 128 + ReLU -> 4 \\
    \midrule
    \multicolumn{2}{l}{\textbf{DNN Property Predictor}} \\
    \midrule
    Block & Specifications \\
    \midrule
    Property Predictor FC & 256 + ReLU -> 128 + ReLU -> 2 \\
    \bottomrule
    \end{tabular}%
  \label{tab:Deepgenerativeparam}%
\end{table}%

\section{Convergence test of the number of sampling points in the latent feature space}\label{sec:appendix3} 
\begin{figure}[!ht]
\centering
\includegraphics[width=0.6\linewidth]{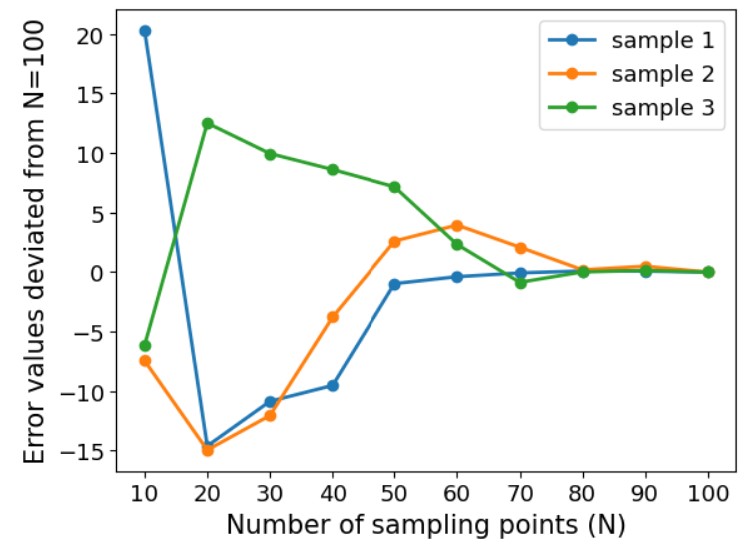}
\caption{\label{fig:A3} Convergence test of the number of sampling points on the latent feature space.}
\end{figure}

The determination of the optimal number of sampling points ($N$) required to accurately estimate the total uncertainty in the latent feature space is achieved through a convergence study. For this purpose, we randomly selected three samples from the validation set and conducted a convergence test by incrementally sampling from the latent feature space, with the number of points ranging from 10 to 100. The specifics of this convergence analysis are documented in Figure \ref{fig:A3}. Based on the results, we settled on $N=80$ as the appropriate number of sampling points within the latent feature space in all our following design cases.

\section{Progressive transfer learning processes of training the proposed deep learning framework}

\begin{table}[htbp]
  \centering
  \caption{Parametric study of different latent dimensions}\label{tab:convergence_study}
    \begin{tabular}{p{12em}cccccc}
    \toprule
    Latent dimension & 4 & 16 & \textbf{32} & 48 & 64 \\
    \midrule
    Recon. MSE training loss & 0.0765 & 0.0117 & \textbf{0.0102} & 0.0105 & 0.009 \\
    \midrule
    Recon. MSE val. loss & 0.1032 & 0.0172 & \textbf{0.0162} & 0.0160 & 0.0152 \\
    \midrule
    Relative error (in $\%$)  & 578.9 & 13.1 & \textbf{6.57} & 5.26 & 0 \\
    \bottomrule
    \end{tabular}%
\end{table}%

\begin{table}[H]
\centering
\caption{Progressive transfer learning-based training strategy for gradually increasing KL loss weights.}
\label{tab:kl_loss_weights}
\begin{tabular}{p{12em}ccccc}
\toprule
Training Iteration & 1 & 2 & 3 & 4 & \textbf{5} \\
\midrule
Reconstruction loss wt. & \multicolumn{5}{c}{\textbf{1}} \\
KL loss wt. (\(\alpha_2\)) & 0 & \(5 \times 10^{-5}\) & \(1 \times 10^{-4}\) & \(5 \times 10^{-4}\) & \bm{$1 \times 10^{-3}$} \\ 
Regression loss wt. (\(\alpha_3\)) & \multicolumn{5}{c}{\textbf{0}} \\
MSE training loss & 0.0102 & 0.0119 & 0.0085 & 0.0077 & \textbf{0.0076} \\ 
MSE val. loss & 0.0162 & 0.0128 & 0.1034 & 0.0098 & \textbf{0.0097} \\
KL training loss & Inf & 8.795 & 4.636 & 3.413 & \textbf{2.791} \\ 
KL val. loss & Inf & 8.799 & 4.636 & 3.414 & \textbf{2.791} \\ 
\midrule
Training Iteration & 6 & 7 & 8 & 9 & 10 \\
\midrule
Reconstruction loss wt. & \multicolumn{5}{c}{1} \\
KL loss wt. (\(\alpha_2\)) & \(5 \times 10^{-3}\) & \(1 \times 10^{-2}\) & \(5 \times 10^{-2}\) & \(1 \times 10^{-1}\) & 1 \\ 
Regression loss wt. (\(\alpha_3\)) & \multicolumn{5}{c}{0} \\
MSE training loss & 0.0085 & 0.0107 & 0.0198 & 0.0328 & 0.1217 \\ 
MSE val. loss & 0.0103 & 0.0120 & 0.0199 & 0.0332 & 0.1222 \\ 
KL training loss & 1.355 & 0.969 & 0.485 & 0.290 & 0.0115 \\ 
KL val. loss & 1.356 & 0.970 & 0.486 & 0.292 & 0.0115 \\ 
\bottomrule
\end{tabular}
\end{table}

\begin{table}[H]
\centering
\caption{Progressive transfer learning-based training strategy for gradually increasing regression loss weights.}
\label{tab:reg_loss_weights}
\resizebox{\textwidth}{!}{
\begin{tabular}{p{10em}ccccccc}
\toprule
Reconstruction loss wt. & \multicolumn{7}{c}{\textbf{1}} \\
\midrule
KL loss wt. (\(\alpha_2\)) & \multicolumn{7}{c}{\bm{$1 \times 10^{-3}$}} \\
\midrule
Regression loss wt. (\(\alpha_3\)) & 0 & \(1 \times 10^{-5}\) & \(1 \times 10^{-4}\) &  \bm{$1 \times 10^{-3}$}  & \(1 \times 10^{-2}\) & \(1 \times 10^{-1}\) & 1 \\
\midrule
Recon. MSE training loss & 0.0076 & 0.0088 & 0.0089 & \textbf{0.0089} & 0.0110 & 0.0163 & 0.02114 \\
\midrule
Recon. MSE val. loss & 0.0097 & 0.0099 & 0.0101 & \textbf{0.0105} & 0.0117 & 0.0171 & 0.02211 \\
\midrule
KL training loss & 2.791 & 2.653 & 2.663 & \textbf{2.686} & 3.133 & 3.825 & 4.273 \\
\midrule
KL val. loss & 2.791 & 2.663 & 2.671 & \textbf{2.594} & 3.135 & 3.826 & 4.268 \\
\midrule
Reg. NLL training loss & 2.831 & -2.646 & -2.892 & \textbf{-3.567} & -3.757 & -3.855 & -7.118 \\
\midrule
Reg. NLL val. loss & 2.839 & -2.585 & -3.074 & \textbf{-2.797} & -3.323 & -3.596 & -3.725 \\
\bottomrule
\end{tabular}
}
\end{table}

\section{Deterministic deep learning framework-based design optimization}\label{sec:appendix1} 
\begin{figure}[!ht]
\centering
\includegraphics[width=1\linewidth]{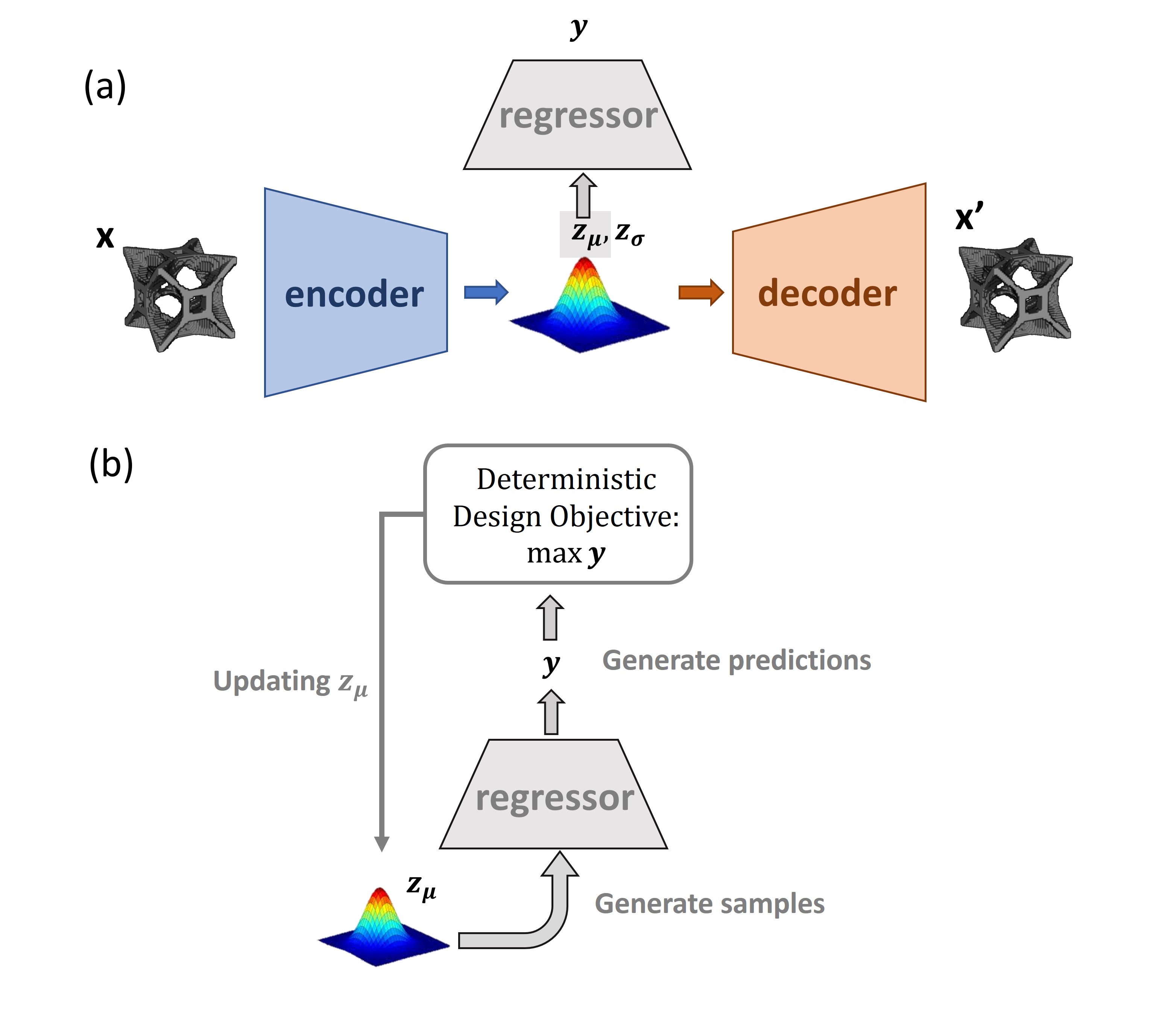}
\caption{\label{fig:A1} (a) A deterministic deep learning model. (b) Deterministic deep learning model-based design approach.}
\end{figure}

We also established a deterministic VAE-based deep learning framework (Figure \ref{fig:A1}a), which comprises an encoder, a decoder, and a feed-forward deep neural network as the property predictor. The hyperparameters of the deterministic deep learning framework are shown in Table \ref{tab:Deepgenerativeparam} in Appendix \ref{sec:appendix2}. To ensure a fair comparison with the uncertainty-aware deep learning framework, we utilize the same training and test set split and the same training process as indicated in Section \ref{subsubsec:training}. We also use the same progressive transfer learning-based training strategy for the model training. Validations of the deterministic deep learning framework's accuracy include voxel-wise comparisons between the original and reconstructed structures, as well as assessing the property predictor's performance in predicting thermal conductivity using Equation \ref{eq:voxelerror} and Equation \ref{eq:R2error}, respectively. The accuracy of the deterministic deep learning framework is presented in Table \ref{tab:A1}.

After successfully training the deterministic VAE-based deep learning framework, which is indicated in Figure \ref{fig:A1}b. The deterministic VAE-based deep learning framework is used in section \ref{subsec:design2}.

\begin{table}[H]
  \centering
  \caption{Reconstruction accuracy of the deep generative model and prediction accuracies of the property predictor.}
  \begin{tabular}{p{8.00em}ccc}
    \toprule
    \multicolumn{1}{c}{\multirow{2}[4]{*}{}} & \multicolumn{1}{c}{\multirow{2}[4]{*}{\makecell{Reconstruction \\Accuracy}}} & \multicolumn{2}{c}{Property} \\
    \cmidrule{3-4}
    \multicolumn{1}{c}{} & & \multicolumn{1}{c}{$E$} & \multicolumn{1}{c}{$\nu$} \\
    \midrule
    training set & 0.9901 & 0.9870 & 0.9214 \\
    validation set & 0.9806 & 0.9846 & 0.9201 \\
    test set & 0.9812 & 0.9855 & 0.9203 \\
    \bottomrule
  \end{tabular}
  \label{tab:A1}
\end{table}

\end{appendices}


\section*{Conflicts of Interest} 
The authors declare no conflict of interest.

\section*{Author Contributions}
Zihan Wang: Methodology, Software, Validation, Formal analysis, Investigation, Writing - original draft, Writing - review \& editing. Anindya Bhaduri: Conceptualization, Methodology, Writing - review \& editing, Supervision. Hongyi Xu: Resources, Software, Methodology, Writing - review. Liping Wang: Writing - review. 



\printbibliography

\end{document}